\documentclass[12pt]{article}
\usepackage{epsfig}
\usepackage{epsf,afterpage}
\def\beq{\begin{eqnarray}}
\def\eeq{\end{eqnarray}}
\def\beqs{\begin{eqnarray*}}
\def\eeqs{\end{eqnarray*}}
\def\dl{\delta}

\newcommand{{\SD}}{\rm SD}

\newcommand{\be}{\begin{equation}}
\newcommand{\ee}{\end{equation}}
\newcommand{\lll}{\langle}
\newcommand{\rrr}{\rangle}

\def\centeron#1#2{{\setbox0=\hbox{#1}\setbox1=\hbox{#2}\ifdim
\wd1>\wd0\kern.5\wd1\kern-.5\wd0\fi
\copy0\kern-.5\wd0\kern-.5\wd1\copy1\ifdim\wd0>\wd1
\kern.5\wd0\kern-.5\wd1\fi}}
\def\ltap{\;\centeron{\raise.35ex\hbox{$<$}}{\lower.65ex\hbox{$\sim$}}\;}
\def\gtap{\;\centeron{\raise.35ex\hbox{$>$}}{\lower.65ex\hbox{$\sim$}}\;}

\begin{document}
\begin{titlepage}
\begin{flushright}
{ITEP-TH-32/04}
\end{flushright}

\vskip 1.2cm

\begin{center}

{\LARGE\bf Current correlators in QCD: OPE versus large distance dynamics. }

\vskip 1.4cm

{\large  V.I.Shevchenko and Yu.A.Simonov}
\\
\vskip 0.3cm
{\it State Research Center \\
Institute of Theoretical and Experimental
Physics, B.Cheremushkinskaya 25 \\ 117218 Moscow, Russia } \\
\vskip 0.25cm e-mail: shevchen,simonov@itep.ru

\vskip 2cm

\begin{abstract}

We analyse the structure of current-current correlators 
in coordinate space in large $N_c$ limit when the corresponding 
spectral density takes the form of an infinite sum over hadron poles.
The latter are computed in the QCD string model with quarks at the ends,
including the lowest states, for all channels. The corresponding 
correlators demonstrate reasonable qualitative agreement with the lattice data
without any additional fits. Different issues concerning the structure of the 
short distance OPE are discussed.

\end{abstract}
\end{center}

\vskip 1.0 cm

\end{titlepage}

\setcounter{footnote}{0} \setcounter{page}{2}
\setcounter{section}{0} \setcounter{subsection}{0}
\setcounter{subsubsection}{0}


\section{ Introduction}

Correlators of hadron currents (denoted as CC in what follows) \be
{\cal P}^{(c)}(x) = \lll j^{(c)} (x) j^{(c)}(0)^{\dagger}\rrr \label{eq1}
\ee are the basic elements in QCD which have been studied in the
framework of perturbation theory \cite{R1} and also using the
Operator Product Expansion (OPE) \cite{R2}. More recently CC have
been the subject of exciting lattice calculations \cite{R3,R4} as
a tool to study nonperturbative vacuum configurations. Since then
it has been realized that CC can provide information of two sorts:
a) one is encoded in the hadron mass poles $m^2_n$ and quark
coupling constants $c_n$ and can be associated with the long
distance dynamics (LDD),  b) another kind of information  can be
extracted from the behavior of ${\cal P}^{(c)}(x)$ at small $x$
(or at large $Q^2$ from the Fourier transform ${\cal P}^{(c)}(Q)$)
with the help of the OPE (see, e.g. \cite{R5}) where the coefficients
represent vacuum matrix elements of local field operators. Since
OPE is assumed to be valid at small distances, one can associate
it with  what we shall call the short distance dynamics (SDD).

There is another way to stress these two different facets of CC
analysis. In the standard Shifman-Vainshtein-Zakharov (SVZ) framework
\cite{R2} one proceeds in two steps. First, the CC is expanded in
local products of fields and, second, the matrix elements of the
latter over (nonperturbative) vacuum states are taken. However, in
principle one can go the other way: first, to take an average of
the CC over the vacuum and, as a second step, to expand the result
in power series at small distances. In the former case the answer
is written in terms of the SVZ vacuum condensates (like $\lll
G_{\mu\nu}^2 \rrr$ or $\lll \bar\psi \psi \rrr$), while in the
latter case one has to deal with the spectral parameters (like
string tension $\sigma$, masses etc.). It is often assumed in the
literature that the results of the two
procedures outlined above must coincide identically. This is an important element 
of the SVZ sum rules: keeping only a few lowest terms in
expansions of {\it both} kinds, one puts them equal
to each other and with the help of some additional mathematical tricks (Borel
transformation) the spectral parameters of low lying
resonances can be expressed in terms of universal vacuum
characteristics like quark and gluon condensates. From the qualitative 
point of view, the approach forbids the so called
$\lambda^2/Q^2$ term (see discussion in \cite{zcn}) since no local gauge-invariant
dimension 2 operator exists in QCD. On the other hand, it is easy
to get nontrivial term of this kind (for example, $\lambda^2 \sim
\sigma$) from the spectral expansion point of view.

The two approaches just described have an important physical
difference. The standard OPE framework encodes the knowledge about
confinement properties of the theory in very indirect way, via
condensates etc. However, nonzero vacuum expectation value of some
local gauge-invariant operator by itself does not imply confinement since 
the
latter is essentially nonlocal phenomenon linked with the
structure of asymptotic states of the theory.\footnote{This is
also true on the lattice where confinement criteria such as area
law or linearly rising potential are physically meaningful only
for times/distances larger than some typical nonperturbative
scale.} On the other hand, the spectral expansion manifestly takes
this aspect into account. One can say that the standard condensate
series is surely sensitive to nonperturbative physics of the
theory but at the same time is "confinement-blind" at the level of
a few lowest terms, since confinement is expected to arise after
summation of the whole series.

In our opinion, there are no solid reasons for the mentioned
coincidence between SDD and LDD to take place beyond perturbation
theory. The key point here is divergence of the series in
question. This precludes to construct one-to-one correspondence
between terms of the small distance expansions based on LDD and on
SDD. In general case each term of the expansion based on LDD is a
result of some SDD subseries summation and vice versa (see examples in \cite{we}).
Moreover, strictly speaking one needs an additional 
prescription to assign physical meaning to the (divergent) sum of
matrix elements of local operators. We do not address general
issues of the OPE structure here, leaving this problem for future
publications (see also \cite{hof} and references therein in this
respect). Instead we shall compare below the explicit  expression
of $CC$ in LDD at large $N_c$ with the lattice data \cite{R3,R4}.
These data provide an essential information on $CC$ in the range
$0\leq x\leq 1.5$ fm where both LDD and SDD are present and
therefore one may check in principle i) the region of validity of
the standard OPE, ii) the transition region from SDD to LDD
and the quark-hadron duality (QHD), iii) the validity of specific
vacuum models and finally iv) the validity of LDD.

In the latter case one has large $N_c$ limit where there is
a solid knowledge of the spectrum properties \cite{R7}. Some 
aspects 
of OPE at large $N_c$ have been discussed recently in \cite{R8}.
Let us remind that the masses and residues (properly normalized) 
of $CC$, $m_n$ and $c_n$, are stable at large
$N_c$ limit, and recent lattice data
(see \cite{R55} and references therein) confirm that typically corrections to stable
quantities for $N_c=4,..,8$ are small.\footnote{In what follows the flavour-singlet channels where
corrections can be large, are not considered.} Moreover at large
$N_c$ one can in principle compute the hadron spectrum (i.e. the
set of $m_n$, $c_n$) quasiclassically in analytic  form for all
$n$ and thus calculate  $CC$ explicitly. This was done
for the masses (see \cite{s1} and references therein) and the residues
(\cite{R10}, see also \cite{R101}) and
compared in \cite{R11} with OPE, SVZ sum rules and experiment. In doing  so one 
encounters and solves several important problems,
which we briefly comment on below.

{First}, the knowledge of $m_n, c_n$ at large $N_c$ allows to find
the limit of the $CC$ at large $Q$ (small $x$) and compare it with
the quark-partonic expression, thus checking the quark-hadron
duality which was suggested long ago and studied since then in 
numerous works (see reviews \cite{R12} containing extensive lists of references
to the original papers). 

Secondly, the large $N_c$ LDD provides the
small $x$ OPE with coefficients depending only on the string
tension $\sigma$, and this LDD OPE is to be distinguished from the
standard, or SDD OPE. These two types  of OPE are in principle
distinct, however in the vector channel both agree with the
experimental $e^+e^-$ data, as was shown in \cite{R11}. Many
efforts  have been undertaken  to match these OPE's (see, e.g.
\cite{RR1,RR2,afonin}),
but here we do not follow this line.

{Thirdly}, one has a unique chance to compare theoretical values
of $CC$ in LDD with experiment and lattice data and discover
its region of validity.

Since theoretical $CC$ in LDD contain no adjustable parameters at
all, and are computed in terms of fixed values of $\sigma$ and
$\alpha_s$ only (and final forms of $CC$ contain only two fixed
masses, expressed through $\sigma, \alpha_s$) this comparison has
a form of a fixed prediction.

The main
purpose of this letter is to construct explicit expressions for
$CC$ in large $N_c$ LDD in the Euclidean 4d $x$-space and to compare those
with lattice results from \cite{R3} and \cite{R4}. This is done in sections 2 and 3. 
The data from \cite{R3} we are going to compare with 
were obtained on $16^3\times 24$ lattice
with the spacing $0.17$ Fm and $\beta=5.7$, while the author of 
\cite{R4} used the lattice of the total size $1.5$ Fm and the spacing $0.13$ Fm at $\beta=5.9$ 
Both simulations were performed in quenched approximation, however the current 
fermions were taken into account differently. An interested reader is encouraged to 
consult the original papers \cite{R3,R4}
for all technical details concerning the simulations. 

In section 3 we calculate the LDD OPE
as the Taylor $x$-expansion and compare it to the standard OPE and
lattice data, defining in this way regions of validity of the
corresponding expansions. There are a few subtle points of
difference between OPE in the coordinate space and in the momentum
space, stated already in the original papers \cite{R2,nsvz}
and discussed also in \cite{chet}. We are not going to repeat old
arguments here and refer an interested reader to the corresponding
literature. The fourth section is devoted to discussion  of the
results and to a short summary of possible perspectives.

\section{The model}

The model of zero-width equidistant resonances \cite{orig,orig2,orig3} has been analyzed in
different respects in connection with OPE and large $N_c$
QCD, where the main interest has been concentrated on the
momentum-space representation \cite{RR1, RR2}. We are following a different path, keeping
ourselves in coordinate space.

As it is discussed in the Introduction, in order to get an
information about nonperturbative  contents of the current
correlators in QCD one usually proceeds along one of two ways. The
first one starts from the small distances - due to asymptotic
freedom it is safe to assume that the leading term at small enough
distances is given by the perturbation theory. There must be power
corrections however, and at this moment the standard method of
SVZ comes into play. The second way goes from the large distances.
Because the theory in question is by assumption confining, at
large distances current-current correlator is dominated by
exchange of the lightest excitation with the particular quantum
numbers. When distance is decreasing, the contribution of higher
states is turning on and at intermediate distances the whole tower
of states plays role. We choose the latter way to go in what
follows. The main object of our interest is the current-current
correlator in coordinate space given by (\ref{eq1}), where the
current $j^{(c)}(x) = \bar\psi(x) \Gamma_c \psi(x)$ is defined for
the given channel $c$. We consider only flavor nonsinglet charged
currents (of the type ${\bar u} \Gamma d$) in this paper. The
matrix $\Gamma_c$ carries Lorentz, flavor and color indices (with
the latter structure being always $\dl_{\alpha\beta}$),
corresponding to quantum numbers of the channel $c$. So we have
for vector ($\rho$ - channel), axial ($a_1$ - channel),
pseudoscalar ($\pi$ - channel) and scalar ($a_0$ - channel) the
following expressions for the (charged) currents\be j^{(v)}_\mu(x)
= {\bar u}\gamma_\mu d \; ; \;\; j^{(a)}_\mu(x) = {\bar
u}\gamma_\mu \gamma_5 d \; ; \;\; j^{(p)}(x) = {\bar u}i \gamma_5
d \; ; \;\; j^{(s)}(x) = {\bar u}d \label{jjj}\ee We adopt the
standard normalization of \cite{R1,R2}, which is different for
neutral currents by a factor $1/\sqrt{2}$ from the one used in
\cite{R5}.

It is convenient to factor out the free part of the correlator
(\ref{eq1}) and to define the ratio $R^{(c)}(x)$ as \be R^{(c)}(x)
= \frac{{\cal P}^{(c)}(x)}{{\cal P}^{(c)}_{free}(x)} \label{R} \ee
where by definition for the vector and axial channels the sum over
Lorentz indices is always taken: \be {\cal P}^{(v,a)}(x) =
g^{\mu\nu}\> {\cal P}^{(v,a)}_{\mu\nu}(x) \label{soi} \ee The free
part ${\cal P}^{(c)}_{free}(x)$ is given by the following
expression in the chiral limit: \be {\cal P}^{(v,a,p,s)}_{free}(x)
= \frac{N_c}{\pi^4 x^6} \cdot (2,\> 2, \> -1, \> -1) \label{free}
\ee
 The important relation $R^{(c)} (x)$ has to obey is given by
\be \lim\limits_{x\to 0} R^{(c)} (x) =1 \label{eq11} \ee Its
physical interpretation depends on what point of view - SDD or LDD
- one takes. In the former case the fact that all effects of
interaction - perturbative and nonperturbative - switches off at
small distances is what is commonly known as asymptotic freedom.
The interpretation is more subtle in the latter case - the fact
that the whole tower of hadron states gives total contribution
coinciding with that of the free quarks (at small distances) is
one of manifestations of the quark-hadron duality. From the point
of view of the spectral expansion, duality is a highly nontrivial
phenomenon since it puts severe bounds on the properties of $\{
m_n \}$, $\{ c_n \}$ at $n\to\infty$. In principle, one should be
able {\it to derive}, rather than {\it postulate} quark-hadron
duality for each given channel starting, for example, from
(relativistic) Hamiltonian and computing its spectrum in this
channel. This problem is beyond the scope of our paper and will be
considered elsewhere. Instead
we will always assume quark-hadron duality in what follows and fix
the coefficients $\{ c_n \}$ to yield duality. In other words, the
condition (\ref{eq11}) holds by definition in our approach.

Another problem is the perturbative corrections. As a matter of principle, 
the LDD procedure remains the same, but one is to include in the spectral 
sum hybrid states. The fact that diagrams of perturbative corrections in confining
vacuum are connected to the hybrid excitations was discovered in \cite{simhyb}
and discussed in \cite{R11}, see also \cite{simyaf,hyb}. At small 
distances $n$-gluon hybrid excitation would 
correspond to the perturbative correction $O(\alpha_s^n)$ from the SDD point of view. 
It is an interesting open problem to establish this relation in explicit 
way and we leave it beyond 
the scope of this paper. The lattice data we are mainly focused on 
provide no clear indication of such corrections as well as of any effects caused 
by perturbative running.\footnote{For example,
due to asymptotic freedom $R^{(c)} (x)$ should approach unity logarithmically slow
when $x\to 0$, which is not seen on the lattice, as figures 1.1 - 4.1 clearly demonstrate.} 
The same reasoning is applicable to anomalous dimensions for nonconserved currents
we take no care of in what follows.

We find it instructive to explain our attitude, 
which in some respects is different from that of the cited papers \cite{RR1,RR2,afonin}.
Our basic assumption is that QCD
exhibits confinement in the form of minimal area law in large 
$N_c$ limit. Since we always have in mind large $N_c$, the picture
of zero-width states is justified and the confinement property
guarantees the spectrum to be discrete. On the other hand, thanks
to the area law one is able to solve the corresponding Hamiltonian
problem for two masses (quarks) connected by the minimal string
\cite{dks} and to get equidistant spectrum (at least
quasiclassically), as it was done in \cite{s1}. 
 Therefore for each channel we are dealing with
the corresponding set of poles $\{ m_n \}$ and residues $\{ c_n
\}$ which in a sense completely characterizes the theory. We get
for the imaginary part of polarization operator \be \frac{1}{\pi}
\> {\mbox{Im}} \>\Pi (s) = \sum\limits_{n=0}^{\infty} c_n
\cdot\delta(s-m_n^2) \label{imp} \ee This is the basic expression
we are to work with below.

The mass spectrum is obtained quasiclassically to have the following form 
\cite{R11,dks,s16}
\be m_n^2 = m_0^2
+ m^2 n \label{rd} \ee where the quantity $m^2$ is defined
universally for all channels to be $m^2 = 4\pi\sigma$ as it was
found by the quasiclassical analysis (see also \cite{lisb} and references therein). 
Here $\sigma$ is
physical string tension.  The residues for the ground state and
the excited states are treated differently. As for the latter,
they are fixed by the requirement of quark-hadron duality. The
lowest state residue is chosen in different physically motivated
ways for different channels (see below). Our approach is
phenomenological in this respect, since we have not computed all
$c_n$ starting from some consistent theoretical scheme (it will be
done elsewhere). It is worth stressing however, that we do
not perform any kind of fitting procedure. In some sense, we have
no fitting parameters at all in our formulas since all quantities
like $m_0$, $\sigma$ take their physical values. We
do not try to fit the lattice data,
instead, we use them for comparison with the results of quasiclassical spectrum
of large $N_c$ QCD.

\subsection{Vector channel}

For conserved vector current one can write the following
expression in momentum space \be {\cal P}^{(v)}_{\mu\nu}(q) =
i\int d^4x \> {\cal P}^{(v)}_{\mu\nu}(x)\exp(iqx) = (q_\mu q_\nu -
q^2 g_{\mu\nu})\Pi^{(v)} (q^2) \label{P} \ee and the determination
of $\Pi^{(v)}(q^2)$ for $-q^2 = Q^2 >0$ is of interest.

It was shown in \cite{R10,R11,dks} that for a system made of relativistic
quarks connected by the straight-line string with tension $\sigma$
one has \be m_n^2 = 2\pi\sigma (2n_r + L) + m_0^2
\;\;\; ; \;\;\; c_n = \frac{N_c}{12\pi^2} \cdot 4\pi\sigma
\label{spektr} \ee which leads to \be \Pi^{(v)}(-Q^2) =  \frac{\lambda_\rho^2}{Q^2 + m_0^2} +
\frac{N_c}{12\pi^2} \> 
\sum\limits_{n=1}^{\infty} \frac{a_n}{Q^2 + m_n^2}
\label{pv} \ee where \be m_n^2 = m^2 n + m_0^2 \;\;\; ; \;\;\; 
a_n = m^2 = 4\pi\sigma \label{resi} \ee 
Despite we work in the large $N_c$ framework, we have kept 
the factor $N_c$ in front of the second term in (\ref{pv}) in order to 
make contact with the asymptotic expression (\ref{dd}).  
The residue of the first ($\rho$-meson) pole
should in principle be calculated in the same dynamical framework, which is used
for the spectrum calculation.\footnote{This is also true for higher resonances, however, we 
choose another way and fix those 
residues by quark-hadron duality for those channels where $c_n$ have not yet been calculated
dynamically. Notice that in some channels like the vector one, dynamically calculated $c_n$ 
indeed provide quark-hadron duality (see \cite{lisb}).}
The condition (\ref{spektr}) provides the quark-hadron duality in
this channel:
 \be
\Pi^{(v)}(-Q^2) \stackrel{Q^2\to\infty}{\longrightarrow} -
\frac{N_c}{12\pi^2} \log\left( \frac{Q^2}{\mu^2}\right) \label{dd}
\ee
Numerically we use values $m_0^2 = M_{\rho}^2 = 0.6$
${\mbox{GeV}}^2$, $\lambda_\rho^2 = 0.047$ ${\mbox{GeV}}^2$ and $m^2 = 4\pi\sigma = 2.1$
${\mbox{GeV}}^2$, corresponding to $\sigma = 0.17$
${\mbox{GeV}}^2$. The value of $\lambda_\rho^2$ is consistent with the one used in \cite{R5,afonin}
and also the one computed in \cite{sim6}. 
Notice that in the latter case we are not to take
into account the large $O(\alpha_s)$ correction which is not seen in lattice simulations. 
It is interesting that the value of the lowest state residue $\lambda_\rho^2$ 
is different from the asymptotic value $c_n$ by less than 15\% in our case.
  
Since we want to compare our model with the lattice results, the Wick 
rotation has to be performed. See Appendix B where all relevant formulas are collected.
The resulting expression in coordinate space, corresponding to
(\ref{pv}) is given by:
\be
R^{(v)}(z) = \xi z^5 K_1(z) + \frac{b z^6}{2^7} \> {\bar p}_b(z) 
\label{vectr}
\ee
where dimensionless distance $z=x_Em_0$ and mass ratio $b={m^2}/{m_0^2}$ have
been introduced and
\be
\xi = \frac{\pi^2}{8} \left( \frac{\lambda_\rho^2}{m_0^2} - \frac{b}{4 
\pi^2}  \right)
\ee
The universal function ${\bar p}_b(z)$ is given by 
\be
{\bar p}_b(z) = \int\limits_0^{\infty}\frac{du}{u^2}\exp\left(-\frac{z^2}{4u}
- u \right)\frac{1+\exp(-bu)(b-1)}{(1-\exp(-bu))^2}
\label{pbbar1}
\ee
As it has already been mentioned, we have three parameters in the expression (\ref{vectr}):
$m_0$, which fixes the overall scale of distance, $\lambda_\rho^2$ and  $b$  
characterizing the spectrum, and all of them are fixed by their physical values.
The resulting plot is shown on Figs. 1.1 and 1.2, to be compared 
with the lattice data of \cite{R3} and \cite{R4}, respectively.
One sees reasonable qualitative agreement with the lattice data. In fact, 
the results from \cite{R3} 
and from \cite{R4} were obtained for different lattice parameters (see the cited papers for details) 
and strictly speaking they do not agree with each other at large distances. However the qualitative
agreement takes the place, and both sets of data reasonably correspond to our 
simple expression (\ref{vectr}).

Of separate interest is the short distance limit of our results.
It is straightforward to expand (\ref{vectr}) near $z=0$, the answer is (see Appendix B) 
$$
R^{(v)}(z) =  \left[  \xi z^4 + \frac{\xi}{4} \> z^6 \log z^2 + ...\right] +
$$
\be
+ \left[1 + \frac{1}{3\cdot 2^7}\left(6b - 6 - b^2 \right) \> {z^4}
+ \frac{3b -2 -b^2}{3\cdot2^8} \> z^6 \log z^2 + ... \right]
\label{shortr}
\ee
The first and the second brackets in the r.h.s. of (\ref{shortr}) correspond
to the first and the second terms in the r.h.s. of (\ref{vectr}), respectively.
The dots stay for higher order terms.
  
Notice an absence of a term proportional to $z^2$ in (\ref{shortr}). It is easy to see by dimensional reasons
that such term would correspond to $\lambda^2 / Q^2$ contribution in 
the polarization operator 
(\ref{pv}). Indeed, high momentum asymptotics of (\ref{pv}) reads:
$$
\Pi^{(v)}(-Q^2) =
\frac{N_c}{12\pi^2} \left[ - \log\left(\frac{Q^2}{m^2}\right) 
+ \left({4 \pi^2 } \lambda_\rho^2 - 
\frac12 {m^2} - m_0^2 \right) 
\cdot \frac{1}{Q^2} + \right.
$$
\be
+ \left.  \left(\frac12 {m_0^4} + \frac12 {m^2 m_0^2} 
+ \frac{1}{12} {m^4} - {4 \pi^2} \lambda_\rho^2 m_0^2  
\right) \cdot  \frac{1}{Q^4} + O\left(Q^{-6}\right)  
\right]
\label{hm}
\ee
so it generally contains $1/Q^2$ term (and in this sense violate the standard OPE where such term 
is absent). In coordinate space, however, this term being multiplied by $Q^2$ produces delta-function like
contribution which we systematically disregard in this paper. In this sense it is impossible to 
detect $1/Q^2$ term by simulation of the correlator in coordinate space (it was noticed in \cite{hints}).  
 However, one may take a different attitude
and put the requirement $\lambda^2 = 0$ right on the 
momentum space expression (\ref{hm}). This would lead to a certain relation between different 
parameters of the model, in our case it is $64 \xi = 2-b$. If one takes all residues
equal (i.e. with $\xi = 0$), it changes to the standard value $b=2$. It is interesting to notice that for the 
parameters we have actually chosen $\lambda^2$ defined by
\be
- \frac{\alpha_s}{4\pi^3} \lambda^2 = \lambda_\rho^2 - \frac{1}{4\pi^2} 
\left(
\frac12 {m^2} + m_0^2 \right) 
\label{lll}
\ee
is equal to $\frac{\alpha_s}{\pi} \lambda^2 = -0.2$ ${\mbox{GeV}}^2$ which is still "tachyonic", to be compared 
with $\frac{\alpha_s}{\pi} \lambda^2 = -(0.07 - 0.09)$ ${\mbox{GeV}}^2$ from \cite{zcn}.

For the sake of completeness let us also cite the standard OPE answer
for the discussed correlator (for the physical value $N_c=3$) \cite{R5,hints}:
\be
R^{(v)}(\tau) = 1+\frac{\alpha_s(\tau)}{\pi} - \frac{\lll ( g G_{\mu\nu}^a)^2 \rrr}{3\cdot 2^7} \> \tau^4 -
\frac{7 \pi^3}{81} \alpha_s \lll {\bar q} q \rrr^2 \> \tau^6 \log  \tau^2 \mu_v^2 + ...
\label{opev}
\ee
where $\tau$ is Euclidean time coinciding with $x_E$ in our case. We use the standard values
\be
\lll ( g G_{\mu\nu}^a)^2 \rrr = 0.5 {\mbox{GeV}}^4 \;\; ; \;\;  
|\lll {\bar q} q \rrr| = (250 {\mbox{MeV}})^3 \label{opiu}
\ee
It is worth noticing\footnote{The authors are indebted to Prof. K.G.Chetyrkin for discussion of this point.}
that the expression (\ref{opev}) contains a contact term $\lll j^\mu j_\nu \rrr$ which is not directly 
seen in the momentum space. This term's contribution is proportional to 
the sixth power of 
$\tau$ in (\ref{opev}), on the other hand, there is no $\alpha_s$ in front of it \cite{chet}. 
Numerically according to \cite{chet}
$$
\lll \bar u \gamma_\mu d \bar d \gamma^\mu u \rrr / \lll {\bar q} q\rrr^2 
= - \frac{1}{3}(0.90\pm 0.15)
$$
We take this circumstance into account by means of numerical redefining 
$\mu_v^2$ in (\ref{opev}) accordingly, while 
still keeping the expression in the form (\ref{opev}). In fact, our main concern is to compare 
(\ref{vectr}) with the lattice, while we need (\ref{opev}) mostly for illustrative purposes.

On Fig. 1.3 we compare the exact expression (\ref{vectr})
with its own short distance expansion (\ref{shortr}) and also with the conventional OPE 
result (\ref{opev}) where the three power terms have been kept in both cases. 
The Figure 1.3 shows striking difference from the Figs. 1.1, 1.2. Sizeable deviations of the standard
OPE from (\ref{vectr}) start as early as at $x\approx 0.4$ Fm. We have also plotted the first 
resonance contribution the answer should converge to at large distances. The sum over resonances
smoothly interpolates between small and large distance regions.

\subsection{Axial channel}

We are now to consider the axial channel. Some care is needed here since
one has to extract the pion contribution. There are two Lorentz structures
in this channel
\be
{\cal P}^{(a)}_{\mu\nu}(q) = i \int d^4 x {\cal P}^{(a)}_{\mu\nu}(x) \exp(iqx) =
(q_\mu q_\nu - g_{\mu\nu} q^2) \Pi_1(q^2) + q_{\mu} q_{\nu} \Pi_2(q^2)
\label{werp}
\ee
Contracting both sides with the tensor $q_\mu q_\nu$ one gets (see Appendix A):
\be
q_\mu q_\nu {\cal P}^{(a)}_{\mu\nu}(q) = q^4 \Pi_2(q^2) = 2(m_u + m_d) \lll {\bar u} u  +
 {\bar d} d \rrr  + (m_u + m_d)^2 \>{\bar{\cal P}}^{(p)}(q^2)
\label{ynd}
\ee

We have put the bar over  ${\cal P}^{(p)}(q^2) $ in order to stress that we
 keep only the pion pole contribution in the pseudoscalar correlator in (\ref{ynd})
(compare with the whole tower taken into account in (\ref{pps})).
This is consistent with the use of Gell-Mann -- Oakes -- Renner
relation in the standard form and such assumption has
approximately the same level of accuracy as this relation itself.
So we have \be {\bar{\cal P}}^{(p)}(-Q^2) =
\frac{\lambda_\pi^2}{Q^2 + m_\pi^2} 
\label{er4} \ee where as usual
$Q^2 = - q^2$. 
The pion residue is fixed by PCAC (see Appendix A)
as \be \lambda_\pi^2 = \frac{2 f_\pi^2 m_\pi^4}{(m_u + m_d)^2}
\label{lambda} \ee where we take $f_\pi = 93$ ${\mbox{MeV}}$,
$m_\pi = 140$ ${\mbox{MeV}}$,  $m_u + m_d = 11$ ${\mbox{MeV}}$, so
that $\lambda_\pi^2 = 0.05$ ${\mbox{GeV}}^4$. 

In principle, one can address in the same framework the question about
corrections to (\ref{testr}) (and correspondingly to (\ref{fgh}),
(\ref{gor2})) due to higher (nonchiral) or multipion states. This
analysis is beyond the scope of the present paper.
Notice the overall plus sign in (\ref{er4}) fixed by
the negative sign of the quark condensate, see Appendix A. One gets the
following expression for $\Pi_2(q^2)$:
\be
q^2 \Pi_2(q^2) = \frac{2f_\pi^2 m_\pi^2}{m_\pi^2 - q^2}
\ee
while for $\Pi_1(q^2)$ we have an expression analogous to (\ref{pv}):
\be
\Pi_1(-Q^2) =
 \frac{N_c}{12\pi^2} \>  \sum\limits_{n=0}^{\infty} \frac{a_n}{Q^2 + m_n^2}  \label{pa} \ee
where we take $m_0 = M_{a_1} = 1.26$ ${\mbox{GeV}}$ while the
residues coincide with those for the vector channel given by
(\ref{resi}). We do not separate the lowest state ($a_1$ resonance) 
as we did for the vector channel since in this correlator 
the role of such state is played by pion. Moreover, as it has already been noticed
the asymptotic value for residues is natural to expect to be 
rather close to the one for the lowest state.

Contraction
of Lorentz indices in (\ref{werp}) leads to \be {\cal
P}^{(a)}(q^2) = g_{\mu\nu} {\cal P}^{(a)}_{\mu\nu}(q) = - 3 q^2
\Pi_1(q^2) + q^2 \Pi_2(q^2) \ee We have therefore \be {\cal
P}^{(a)}(-Q^2) = \frac{2f_\pi^2 m_\pi^2}{m_\pi^2 + Q^2} + 3Q^2
\frac{N_c}{12\pi^2} \> \sum\limits_{n=0}^{\infty} \frac{a_n}{Q^2 +
m_n^2} \label{paa} \ee
Fourier thansformation to the coordinate space is straightforward, the answer is 
given by the following expression:
\be
R^{(a)}(z) = -\frac{\pi^2}{12}\> f^2 \gamma^3 z^5 K_1(\gamma z) +
\frac{b z^6}{2^7} \> {\bar p}_b(z) 
\label{raxial}
\ee
The following notations have been introduced:
$\gamma = {m_\pi}/{m_0}$, $ f = f_\pi / m_0$. Notice the different sign of pion contribution 
and that of the tower. The correlator becomes negative at large distances.
On Figs. 2.1 and 2.2 expression (\ref{raxial}) is compared with the lattice data from \cite{R3} and from
\cite{R4}.
The short distance expansion of (\ref{raxial}) is given by 
$$
R^{(a)}(z) =  -\frac{\pi^2}{12}\> f^2 \gamma^3 \left[ \frac{1}{\gamma}\> 
z^4 + 
\frac{\gamma}{4} z^6 \log (\gamma z)^2 + ...\right] +
$$
\be
+ \left[1 + \frac{1}{3\cdot 2^7}\left(6b - 6 - b^2 \right) \> {z^4}
+ \frac{3b -2 -b^2}{3\cdot2^8} \> z^6 \log z^2 + ... \right]
\label{shorta}
\ee
while the standard OPE series can be written as 
\be
R^{(a)}(\tau) = 1+\frac{\alpha_s(\tau)}{\pi} - \frac{\lll ( g G_{\mu\nu}^a)^2 \rrr}{3\cdot 2^7} \> \tau^4 +
\frac{11 \pi^3}{81} \alpha_s \lll {\bar q} q \rrr^2 \> \tau^6 \log  \tau^2 \mu_a^2 + ...
\label{opea}
\ee
where we again denote $x_E$ as $\tau$ and the discussion we had after (\ref{opiu}) 
is also valid here. 
The comparison of (\ref{raxial}), (\ref{shorta}) and (\ref{opea})
is presented on Fig. 2.3. One can see that the standard OPE (as well as the short distance 
expansion (\ref{shorta})) deviates significantly from the lattice 
data at distances larger than 0.4 Fm, while the full spectral density (\ref{raxial}) reproduces 
the latter smoothly.

\subsection{Pseudoscalar channel}

This channel contains the light pion which makes it different from
the others. The corresponding Hamiltonian analysis should take the
chiral nature of pion into account. The resulting spectrum
contains the lightest excitation (pion) which is massless in the
chiral limit plus a tower of massive states with the masses
experiencing  chiral shifts, which decrease with $n$ (see
\cite{puz}). Therefore we proceed as before, separating the lowest
state (pion) and fixing the residues of all other poles by the
duality requirement. The correlator reads as \be {\cal P}^{(p)}(q)
= i \int d^4 x \> \lll 0 | {\mbox{T}}\> j^{(p)}(x)
j^{(p)}(y)^\dagger | 0 \rrr\>\exp(iq(x-y)) \label{eitps} \ee with
$j^{(p)}(x) = {\bar u}(x)i \gamma_5 d(x) $. We take the
polarization operator in the following form: \be {\cal
P}^{(p)}(-Q^2) = \frac{\lambda_\pi^2}{Q^2 + m_\pi^2} +
\sum\limits_{n=1}^{\infty} \frac{a_n^{(p)}}{Q^2 + m_n^2}
\label{pps} \ee We neglect the
chiral shifts for higher states in what follows and take the
spectrum of the tower in the form: \be m_n^2 = m_1^2 + m^2(n-1)\;\; ; \;\; n = 1, 2,... \ee
where $m_1 = 1.3$ ${\mbox{GeV}}$ as for $\pi(1300)$ state. The
residues $a_n^{(p)}$ should depend on $n$ in such a way that the
leading term is linear: \be a_n^{(p)} = a^{(p)} n + {\bar a}^{(p)}
+  {\cal O} \left({n}^{-1}\right) \label{lasd} \ee This form is
dictated by the quark-hadron duality, postulated throughout the
paper. However, duality requirement allows to fix only the leading
term, $a^{(p)}$, but not the subleading one  ${\bar a}^{(p)}$
which plays a role of a free parameter in our analysis. For the
coefficient $a^{(p)}$ we obtain \be a^{(p)} = \frac{N_c}{8\pi^2}
\> m^4 \ee which provides \be {\cal P}^{(p)}(-Q^2) =
\stackrel{Q^2\to\infty}{\longrightarrow}  \frac{N_c}{8\pi^2} Q^2
\log\left( \frac{Q^2}{\mu^2}\right) \label{dd1} \ee
The full expression for the correlator's ratio in coordinate space has the form:
\be
{R}^{(p)}(z) = \frac{\pi^2}{12} L^2 \gamma z^5 K_1(\gamma z) + \frac{b^2 
(a+1) - b}{2^5} 
z^4 \>{\hat p}_b(z)
+  \frac{b z^6}{2^7} \> {\bar p}_b(z) 
\label{pseudosc}
\ee
where we use the following notation:
\be
\gamma = \frac{m_\pi}{m_1} \;\; ; \;\; L^2 = \frac{\lambda_\pi^2}{m_1^4} \;\; ; \;\;
a= \frac{8\pi^2 {\bar a}_p}{3 m^4} \;\; ; \;\; b = \frac{m^2}{m_1^2}\;\; ; 
\;\; z = m_1 x_E
\ee
Notice the relative positive sign of the first and the third terms in the rhs of (\ref{pseudosc}).
The parameters $\gamma$, $L$ and $b$ are fixed by their physical values while the parameter
$a$ is free in our approach, on physical grounds it is reasonable to 
expect it to be of the order of unity. On  
Figure 3.1 and 3.2 we plot the function $ {R}^{(p)}(z) $ together
with the lattice data for the choice $a=0$. 
The Figure 3.4 shows the curves (\ref{pseudosc})
for the choices $a=-1,0,2$. Varying $a$ slightly 
affects intermediate distance behavior of the correlator but does not 
touch short and long distance asymptotics.

As above, we consider the short distance expansions. For (\ref{pseudosc}) it reads
$$
R^{(p)}(z) =  \frac{\pi^2}{12}\> L^2 \gamma \left[ \frac{1}{\gamma}\> z^4 
+ 
\frac{\gamma}{4} z^6 \log (\gamma z)^2 + ...\right] +
\left[ \frac{b (a+1) - 1}{2^3} \> z^2 + ... \right] +
$$
\be
+ \left[1 + \frac{1}{3\cdot 2^7}\left(6b - 6 - b^2 \right) \> {z^4}
+ \frac{3b -2 -b^2}{3\cdot2^8} \> z^6 \log z^2 + ... \right]
\label{shortps}
\ee
where dots stay for higher powers of $z^2$. In contrast to (\ref{shortr}), (\ref{shorta})
all non-negative powers of $z^2$ are present in (\ref{shortps}).
The conventional OPE answer in this channel has the form (see, e.g. \cite{hints}):
\be
R^{(p)}(\tau) = 1 + \frac{\lll ( g G_{\mu\nu}^a)^2 \rrr}{3\cdot 2^7} \> \tau^4 -
\frac{7 \pi^3}{81} \alpha_s \lll {\bar q} q \rrr^2 \> \tau^6 \log  \tau^2 \mu_p^2 + ...
\label{opeps}
\ee
If one includes the quadratic power correction in the OPE expansion (\ref{opeps}), the corresponding 
relation is to be
\be
\frac{\alpha_s}{\pi} \lambda^2 = - \frac{b (a+1) - 1}{4} 
\label{polk}
\ee
Since $a$ is a free parameter in our approach we can make no reasonable estimate of $\lambda^2$
from (\ref{polk}). 
The other way around, fixing the l.h.s. of (\ref{polk}) by (\ref{lll}) corresponds to 
the value $a = 0.4$.  

The Figure 3.3 shows the exact answer (\ref{pseudosc}) together with the short distance asymptotics
(\ref{shortps}) and (\ref{opeps}). Analogously to the other channels, short-distance expansions break down
at $x\approx 0.4$ Fm.

\subsection{Scalar channel}

This channel is presumably the most complex one since it requires some
understanding of the $a_0$ meson nature,\footnote{Let us remind that we work with $I=1$
case only, so we have no mixing with the pure glue states and do not 
consider $f_0$ meson.} 
in particular, the complicated dynamical problem of the role played by the four
quark state admixture in these states. However, we do not
address this set of questions here since the lattice data of
\cite{R3} we are going to compare with are obtained in quenched
approximation. We also take the set of data corresponding to the
largest value of current quark mass used in simulations \cite{R4}.
We also have in mind large $N_c$ limit, where four quark admixture
vanishes. We write therefore \be
  {\cal P}^{(s)}(-Q^2) =  \sum\limits_{n=0}^{\infty}
\frac{a_n^{(s)}}{Q^2 + m_n^2}
\label{pss}
\ee
with the corresponding quark-hadron duality condition
\be
{\cal P}^{(s)}(-Q^2) = \stackrel{Q^2\to\infty}{\longrightarrow}
 \frac{N_c}{8\pi^2} Q^2 \log\left( \frac{Q^2}{\mu^2}\right)
\label{dd2} \ee The expression for residues coincides with
(\ref{lasd}): \be a_n^{(s)} = a^{(s)} n + {\bar a}^{(s)} + {\cal
O} \left(\frac{1}{n}\right) \label{lasd2} \ee and 
\be a^{(s)} = \frac{N_c}{8\pi^2}
\> m^4 \ee 
The result for the ratio $R^{(s)}(z)$ is 
\be
{R}^{(s)}(z) = \frac{b (ab-1) }{2^5} 
z^4 \>{\hat p}_b(z) +  \frac{b z^6}{2^7} \> {\bar p}_b(z) 
\label{scal}
\ee
The notation is standard:
$$
a= \frac{8\pi^2 {\bar a}_s}{3 m^4} \;\; ; \;\; b = \frac{m^2}{m_0^2}\;\; ; 
\;\; z = m_0 x_E
$$
Notice the difference between (\ref{scal}) and (\ref{pseudosc}). In the latter case the $\pi$ - meson pole stay 
away from the tower, but the numbering of states still respects the fact that the pion is the lowest, $n=0$ state.
This is in line with \cite{R3}. In the scalar channel the tower itself starts from the lowest state. 
This causes $b^2 (a+1) - b$ factor in (\ref{pseudosc}) instead of $b (ab-1) $ in (\ref{scal}). Also 
the parameter $b$ is defined differently. One may check that 
if mass and residue of the lowest state are chosen in such a way that this state also belongs to the 
equidistant tower, the expression (\ref{scal}) identically coincides with (\ref{pseudosc}), as it should be.

The standard OPE answer is given by \cite{R2}
\be
R^{(s)}(\tau) = 1 + \frac{\lll ( g G_{\mu\nu}^a)^2 \rrr}{3\cdot 2^7} \> \tau^4 +
\frac{15 \pi^3}{81} \alpha_s \lll {\bar q} q \rrr^2 \> \tau^6 \log  \tau^2 \mu_s^2 + ...
\label{opes}
\ee
while the short distance expansion of (\ref{scal}) takes the form
$$
R^{(s)}(z) =  \left[ \frac{b a - 1}{2^3} \> z^2 + ... \right] +
$$
\be
+ \left[1 + \frac{1}{3\cdot 2^7}\left(6b - 6 - b^2 \right) \> {z^4}
+ \frac{3b -2 -b^2}{3\cdot2^8} \> z^6 \log z^2 + ... \right]
\label{shortsc}
\ee

The Figure 4.1 shows the curve defined by (\ref{scal}) for different values of $a$ against the 
data from \cite{R3}. Unfortunately, the data in this channel are not reliable enough, which can 
be clearly seen on the Figure 4.2, where the lattice correlator from \cite{R4} 
becomes negative at $x\approx 0.5$ Fm due to lattice artefacts, in contradiction with 
positivity of the spectral density.\footnote{The authors are grateful to Prof. De Grand for discussion of 
the problems in this channel.} The short distance expansions explode even earlier (see Figure 4.3).

\section{Discussion of the results and conclusion}

We find the results of comparison of the lattice data and the model 
to be quite remarkable. It is worth stressing 
that we have worked with the model of two relativistic
quarks connected by the string with the tension $\sigma$, which is the only dimensionful
parameter of the model. Using this picture, first, the lowest 
resonance masses were computed in all channels \cite{s1} and found to be in reasonable agreement
with their experimental values. Second, the quasiclassical asymptotic for the mass spectrum
was calculated \cite{dks} which has the same pattern for all
channels. The latter must provide exact quark-hadron duality, which was explicitly 
shown in the vector case in \cite{R11,sim6}.
We have plugged these two ingredients into the corresponding spectral 
density and compared the results with the available lattice simulations \cite{R3,R4}.
Another comparison is made for the
standard SVZ short distance OPE expansion and the corresponding expansion provided by 
our spectral density. It is shown that they strongly deviate from the full curve (and from the lattice data)
for the distances larger than $0.35-0.45$ Fm, as one could expect.  
 
The main lesson is that one needs quite a few inputs (correct lowest resonance mass, 
correct lowest resonance residue and correct
asymptotic behavior dictated by quark-hadron duality) to reproduce lattice data in a reasonable 
way. We have taken these inputs from the large $N_c$ model of QCD string with quarks at the ends
\cite{dks}. The absence of precise quantitative agreement between our results and the lattice 
should not disappoint since many effects have been
ignored ($1/N_c$ effects, perturbative exchanges etc), notice also that the two sets of 
lattice data we have used do not agree with each other on quantitative level. 
On the other hand,  qualitative agreement is rather good and certainly better than that
of the standard short distance OPE in coordinate space. Notice that we have not performed
any fitting of the lattice data, the latter was compared with the curves 
(\ref{vectr}),(\ref{raxial}),(\ref{pseudosc}),(\ref{scal}) computed independently. 
If we had fitted the data with these expressions, the agreement would have been much better.

The next logical step is to compare the predictions of large $N_c$ QCD string 
model with the data taken from real experiments, i.e. the data on $\tau$ - decay rates
obtained by ALEPH and OPAL collaborations. 
The accuracy of the data is much higher than what is possible to reach 
on the lattice, and this fact provides a challenging task for theory.
The $\tau$ - decay results have been already addressed in the framework of conventional OPE \cite{ioffe}
 and in the instanton model \cite{sss}. The issues of large $N_c$ limit interpretation of 
the data have also been discussed \cite{eraf}. We are going to analyse these data in our framework 
in a separate publication.

{\bf Acknowledgments }

\bigskip

The authors are grateful to J.W. Negele and T.A. DeGrand for
providing their numerical data and valuable explanations.
The authors wish to thank A.V.Nefediev, V.I.Zakharov, K.G.Chetyrkin and A.A.Pivovarov for useful
discussions. Support from the Federal programme of Russian Ministry of industry,
science and technology 40.052.1.1.1112 and from the grant for scientific schools NS-1774.2003.2 
is acknowledged. V.Sh. is thankful to the non-profit "Dynasty"
foundation and ICFPM for financial support.

\section*{Appendix A}

This Appendix contains a textbook material (see, e.g. \cite{R1}),
however we find it useful for the sake of completeness to rederive
here the equation (\ref{ynd}) in our conventions. The following
expressions for charged axial and pseudoscalar currents have been
introduced: \be j^{(a)}_\mu(x) = {\bar u}(x)\gamma_\mu \gamma_5
d(x) \;\;\; ; \;\;\; j^{(p)}(x) = {\bar u}(x)i \gamma_5 d(x)
\label{poik} \ee Using Dirac equation \be
(i\gamma_\mu(\partial_\mu - ig A_\mu(x)) - m)\psi(x) = 0 \ee one
gets the well known expression for PCAC: \be
\partial_\mu j^{(a)}_\mu(x) = (m_u + m_d) j^{(p)}(x)
\label{pcac}
\ee
Next, one defines the following matrix elements of the currents between vacuum and one pion state
\be
\lll 0 | j^{(a)}_\mu(x) |\pi(p)\rrr = i \sqrt{2}\> f_\pi p_\mu \exp(-ipx) \;\; ; \;\;
\lll 0 | j^{(p)}(x) |\pi(p)\rrr = \lambda_\pi \exp(-ipx)
\label{pik}
\ee
Our normalization coincides with \cite{R1} but $\sqrt{2}$ different from \cite{R5}.
Correspondingly, we have $f_\pi = 93$ $\mbox{MeV}$ instead of $131$ $\mbox{MeV}$ value for $f_\pi$ used
in \cite{R5}.

Combining (\ref{pcac}) and (\ref{pik}) one gets \be \lambda_\pi =
\frac{\sqrt{2}\> f_\pi m_\pi^2}{m_u + m_d} \label{lam2} \ee
Consider now the current-current correlator in momentum space \be
{\cal P}^{(a)}_{\mu\nu}(q) = i \int d^4 x \>\lll 0 | {\mbox{T}}\>
j^{(a)}_\mu(x) j^{(a)}_\nu(y)^\dagger |0\rrr\>\exp(iq(x-y))
\label{eit} \ee Making use of free equal-time anticommutation
relations for the quark fields, it is straightforward to obtain:
\be q_\mu q_\nu {\cal P}^{(a)}_{\mu\nu}(q) = (m_u + m_d)^2 {\cal
P}^{(p)}(q^2)  + 2 (m_u + m_d) \lll {\bar u} u + {\bar d} d \rrr
\label{kgh} \ee where ${\cal P}^{(p)}(q)$ is given by
(\ref{eitps}). Assuming that in the chiral limit one pion state
contribution dominates in the pseudoscalar correlator at $q^2 \to
0$: \be {\cal P}^{(p)}(q^2) \stackrel{q^2 \to 0}{\longrightarrow}
\frac{\lambda_\pi^2}{m_\pi^2} \label{testr} \ee one gets \be 2
m_\pi^2 \> \lll {\bar u} u + {\bar d} d \rrr + (m_u + m_d)
\lambda_\pi^2 = 0 \label{fgh} \ee which together with (\ref{lam2})
is nothing but the Gell-Mann -- Oakes -- Renner relation
\be (m_u + m_d) \lll {\bar u} u + {\bar d} d \rrr = -
f_\pi^2 m_\pi^2 \label{gor2} \ee up to the terms of higher order
in $m_\pi^2$. Taking into account the definition (\ref{werp}) we
come to the expression (\ref{ynd}) used in the main text.

\section*{Appendix B}

We collect here the basic formulas used in the main text.    
The standard metric conventions, correlator definitions, Fourier transforms are 
adopted. The free propagator is given by \be \lll \psi(x) \bar\psi(0) \rrr = i
\int\frac{d^4 q}{(2\pi)^4} \> \frac{ {\hat q} + m}{q^2 - m^2}\>
\exp(-iqx) = \left(i\gamma_\mu \frac{\partial}{\partial x_\mu} +
m\right) \; \frac{m^2}{4\pi^2 z} \> K_1\left(mz\right)\ee 
for space-like $x$, where $z=m\sqrt{-x^2}$.
Four vectors in Minkowskii and Euclidean spaces are defined as follows:
\be
q = (q_0, {\bf q}) \;\; ; \;\; q^2 \equiv q_M^2 = q_0^2 - {\bf q}^2 \;\; ;\;\;
Q^2 \equiv q_E^2 = - q_M^2
\ee
and we often omit the subscripts $E, M$ if it is not misleading. 
Notice that the Wick rotation changes the free propagator sign :
\be {\cal P}_{free}(x) \sim
\frac{1}{(x^2)^3} \;
\to \; - \frac{1}{(x_E^2)^3} 
\label{free2}
\ee
The Fourier transform for the free propagator is given by (see, e.g. \cite{nsvz}):
\be
 \int d^4 x \>\frac{1}{(x^2)^3} \> \exp(iqx) \leftrightarrow i \frac{\pi^2}{8} (-q^2) \log(-q^2)
\ee

Let us calculate the function ${\bar P}(x)$ we have made use of in the main text.
It is defined as follows: 
\be {\bar P}(x) = 3i \int \frac{d^4
q}{(2\pi)^4} \> q^2 \sum\limits_{n=0}^{\infty} \frac{1}{-q^2 + m_n^2} \exp(-iqx) \label{asd} 
\ee 
where $m_n^2 = m_0^2 + m^2 n$. 
The sum
over $n$ is divergent and the divergent ($q^2$ - independent)
constant term is usually eliminated by renormalization $\Pi(q^2)
\to \Pi(q^2) - \Pi(0)$ in momentum space. We proceed in a
different way here, simply interchanging the summation over $n$
and integration over $q$ for $x\neq 0$, having in mind condition
(\ref{eq11}) then $x\to 0$. We thus systematically disregard all terms proportional 
to $\delta(x)$ or derivatives of the delta function. In this way one obtains \be {\bar P}(x) =
- 3 \left(-\frac{\partial^2}{\partial x^2}\right)\;
\sum\limits_{n=0}^{\infty} \frac{m_n}{4\pi^2 \sqrt{-x^2}} \> {{K}}_1\left(m_n\sqrt{-x^2}\right)
\label{K1-0} \ee 
We switch to the Euclidean notations in the summand (\ref{K1-0}), with $x_E = \sqrt{-x^2} > 0$.
Using integral representation for the McDonald
function \be {{K}}_1(w) = \frac{w}{4} \>
\int\limits_0^{\infty} \frac{dt}{t^2} \exp\left( -t - \frac{w^2}{4t} \right) 
\label{K1-1} \ee and performing the summation
over $n$ \be \sum\limits_{n=0}^{\infty} \exp\left(- \frac{x_E^2m_n^2}{4t} 
\right) = \frac{\exp\left(-\frac{x_E^2
m_0^2}{4t}\right)}{1-\exp\left(-\frac{x_E^2 m^2}{4t}\right)}
\label{K1-2} \ee we obtain \be {\bar P}(x) = 
\frac{3}{16 \pi^2} \left(
\frac{\partial^2}{\partial x^2} \right) \int\limits_0^{\infty}
\frac{dt}{t^2} \exp(-t) f_0\left(\frac{x_E^2}{t}\right) \label{k1-3}
\ee where
\be
f_0\left(\frac{x_E^2}{t}\right) = \sum\limits_{n=0}^{\infty} m_n^2
\exp\left(- \frac{x_E^2 m_n^2}{4t} \right)
=
\frac{\exp\left(-\frac{x_E^2
m_0^2}{4t}\right)}{\left(1-\exp\left(-\frac{x_E^2 m^2}{4t}\right)\right)^2} \cdot
\left( m_0^2 + \left(m^2 - m_0^2\right)\exp\left(-\frac{x_E^2
m^2}{4t}\right)    \right)
\ee

To represent the results in more convenient form we use the
notation $z=x_E m_0$ for dimensionless distance and $b=m^2/m_0^2$
for the quanta/gap ratio. Changing integration variables $t\to u$ such that  $4ut = z^2$,
one gets 
\be
{\bar P} (x) = \frac{3}{16 \pi^2}\> \int\limits_0^{\infty}du \left(
\frac{\partial^2}{\partial x^2} \left[
\frac{4}{x_E^2}\exp\left(-\frac{x_E^2m_0^2}{4u}\right)\right]\right)\cdot
\frac{\exp(-u)}{(1-\exp(-bu))^2}\left(1+\exp(-bu)(b-1)\right)
\label{kl-5} \ee 
Performing differentiation, we finally obtain
\be
{\bar P}(x) = - \frac{3m_0^4}{16\pi^2}\> {\bar p}_b(z)
\label{pbar}
\ee
where the universal function ${\bar p}_b(z)$ has the following form
\be
{\bar p}_b(z) = \int\limits_0^{\infty}\frac{du}{u^2}\exp\left(-\frac{z^2}{4u}
- u \right)\frac{1+\exp(-bu)(b-1)}{(1-\exp(-bu))^2}
\label{pbbar}
\ee
This result has been used in the main text.

For scalar and pseudoscalar channels one also needs
the function ${\hat P}(x)$ defined as 
\be
{\hat P}(x) = -i \int \frac{d^4
q}{(2\pi)^4} \>  \sum\limits_{n=0}^{\infty} \frac{1}{-q^2 + m_n^2} \exp(-iqx) \label{asd4} 
\ee 
where again $m_n^2 = m_0^2 + m^2 n$. Performing completely analogous calculations one gets
\be
{\hat p}_b(z) = 4\pi^2 x_E^2 \> {\hat P}(x) = \int\limits_0^{\infty}{du}\> \exp\left(-\frac{z^2}{4u}
- u \right)\frac{1+\exp(-bu)(b-1)}{(1-\exp(-bu))^2}
\label{phat}
\ee

The short-distance asymptotics of ${\bar p}_b(z)$ can be easily found from the definition (\ref{pbbar}), 
it is given by:
\be
{\bar p}_b(z) = \frac{128}{b} \cdot \frac{1}{z^6} + \left(2-\frac{2}{b} - \frac{b}{3} \right) \cdot \frac{1}{z^2}
+ \frac{3b -2 -b^2}{6b} \> \log z^2 + ...
\label{shortd}
\ee
where the dots denote constant ($z^2$ -independent) term and positive powers of $z^2$.

\newpage


\newpage

\begin{center}
{\Large  Figure captions}
\end{center}

\bigskip

\begin{itemize}
\item[\sf
Fig. 1.1]  Lattice data from \cite{R3} for the vector 
channel correlator vs expression (14). 
\item[\sf
Fig. 1.2] Lattice data from \cite{R4} for the vector
channel correlator vs expression (14).
\item[\sf
Fig. 1.3] Expression (14) (solid curve) together with its own short 
distance expansion (17) (dashed-dotted curve) and short distance
OPE expansion (20) (dashed curve). The lightest resonance contribution is 
also shown (double dotted curve).  
\item[\sf
Fig. 2.1]  Lattice data from \cite{R3} for the axial
channel correlator vs expression (30).
\item[\sf
Fig. 2.2] Lattice data from \cite{R4} for the axial
channel correlator vs expression (30).
\item[\sf
Fig. 2.3] Expression (30) (solid curve) together with its own short
distance expansion (31) (dashed-dotted curve) and short distance
OPE expansion (32) (dashed curve).
\item[\sf
Fig. 3.1]  Lattice data from \cite{R3} for the pseudoscalar
channel correlator vs expression (39).
\item[\sf
Fig. 3.2] Lattice data from \cite{R4} for the pseudoscalar
channel correlator vs expression (39).
\item[\sf
Fig. 3.3] Expression (39) (solid curve) together with its own short
distance expansion (41) (dashed-dotted curve) and short distance
OPE expansion (42) (dashed curve).
\item[\sf
Fig. 3.4] Expression (39) for different values of the parameter $a$:
$a=2$ (upper curve), $a=0$ (solid curve) and $a=-1$ (lower curve).
\item[\sf
Fig. 4.1]  Lattice data from \cite{R3} for the scalar
channel correlator vs expression (48) for different values of $a$: $a=2$
(upper curve), $a=1$ (middle curve) and $a=0$ (lower curve).
\item[\sf
Fig. 4.2] Lattice data from \cite{R4} for the scalar
channel correlator vs expression (48).
\item[\sf
Fig. 4.3] Expression (48) (solid curve) together with its own short
distance expansion (50) (dashed-dotted curve) and short distance
OPE expansion (49) (dashed curve).
\end{itemize}

\newpage

\begin{figure}[!htb]
\begin{center}
\begin{tabular}{lr}
\includegraphics[angle=-00,height=82mm,width=82mm,clip=true]{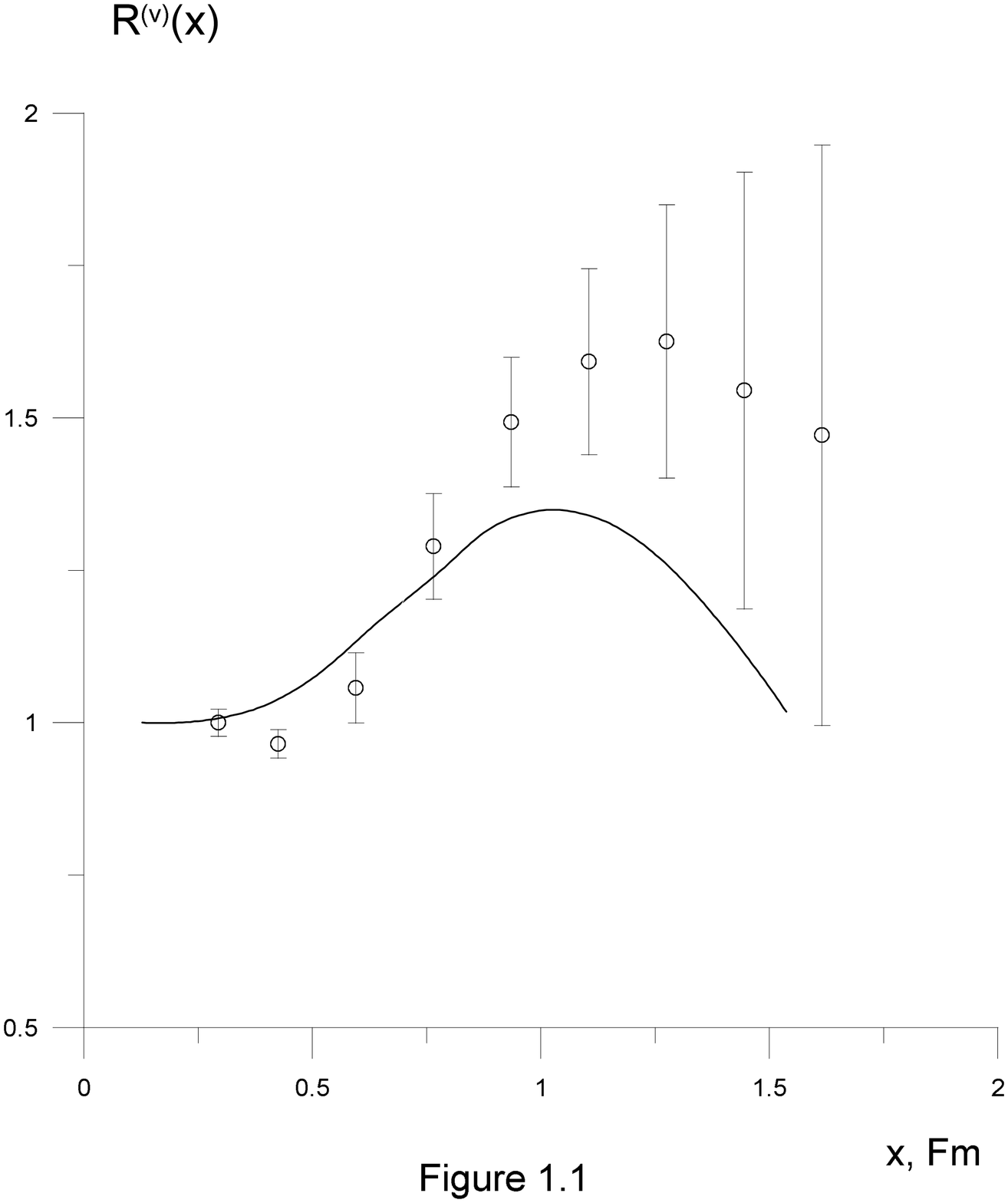} 
&
\includegraphics[angle=-00,height=82mm,width=82mm,clip=true]{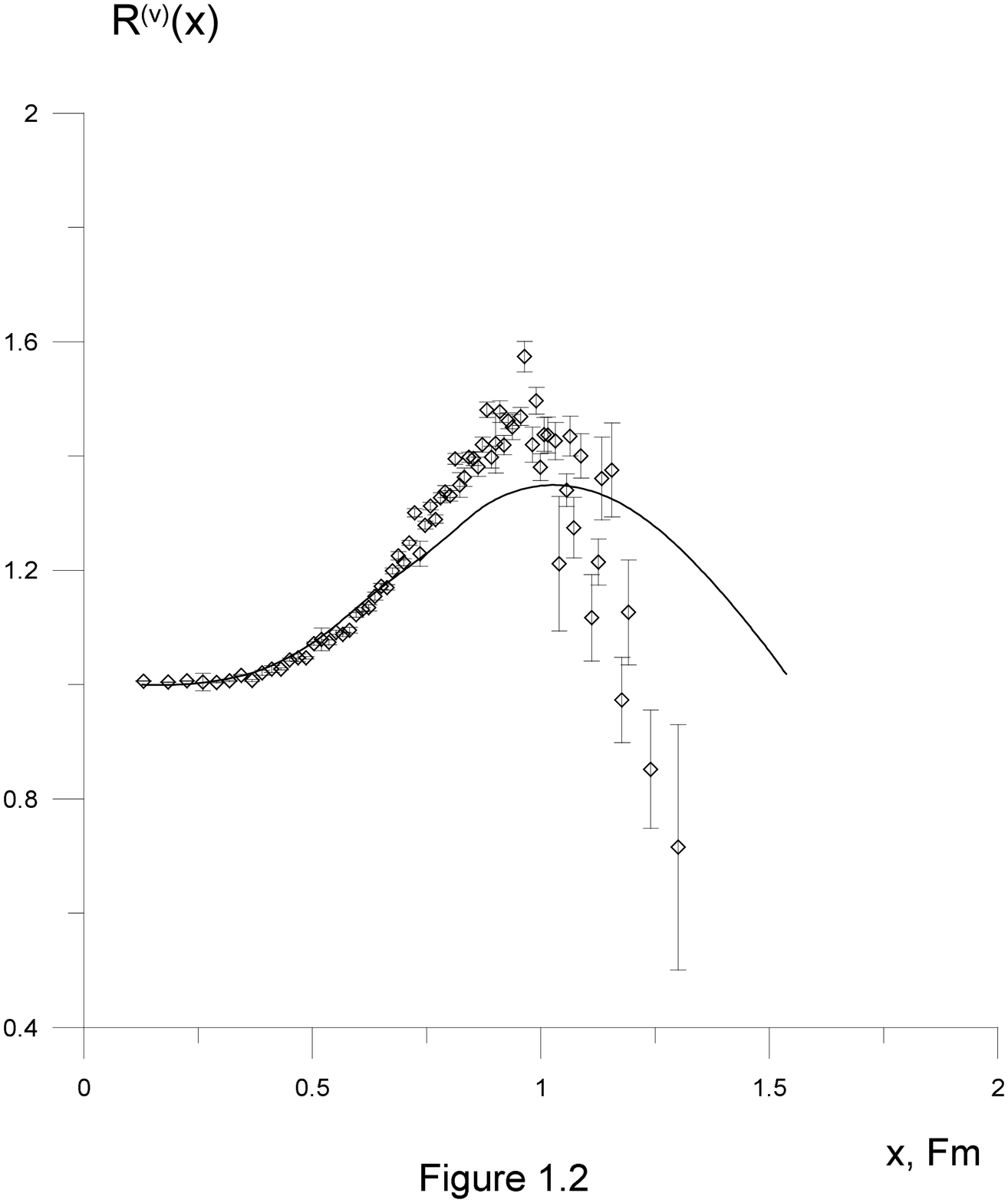} 
\end{tabular}
\end{center}
\vspace{2cm}
\begin{center}
\includegraphics[angle=-00,height=82mm,width=82mm,clip=true]{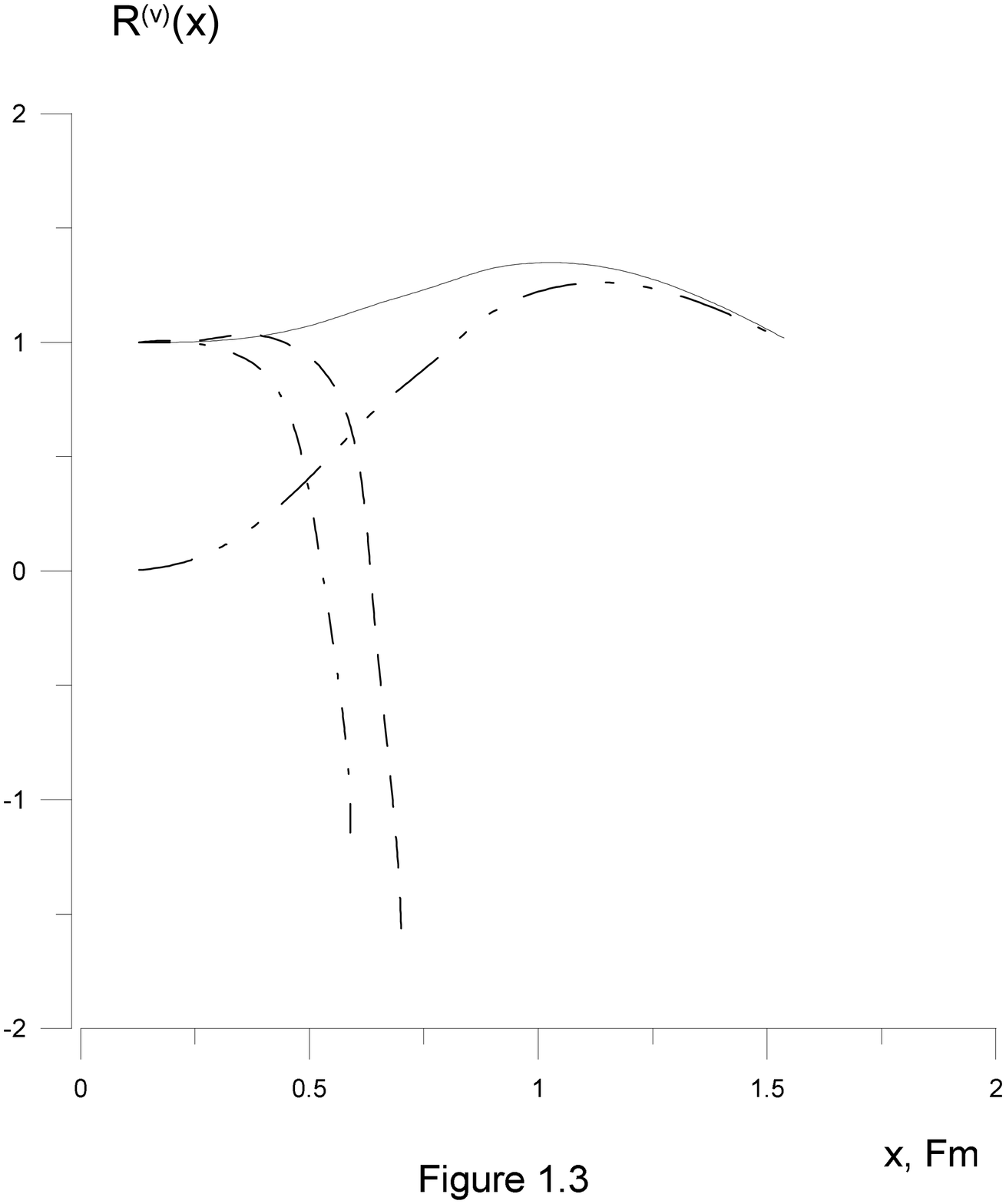} 
\end{center}
\end{figure}

\begin{figure}[!htb]
\begin{center}
\begin{tabular}{lr}
\includegraphics[angle=-00,height=82mm,width=82mm,clip=true]{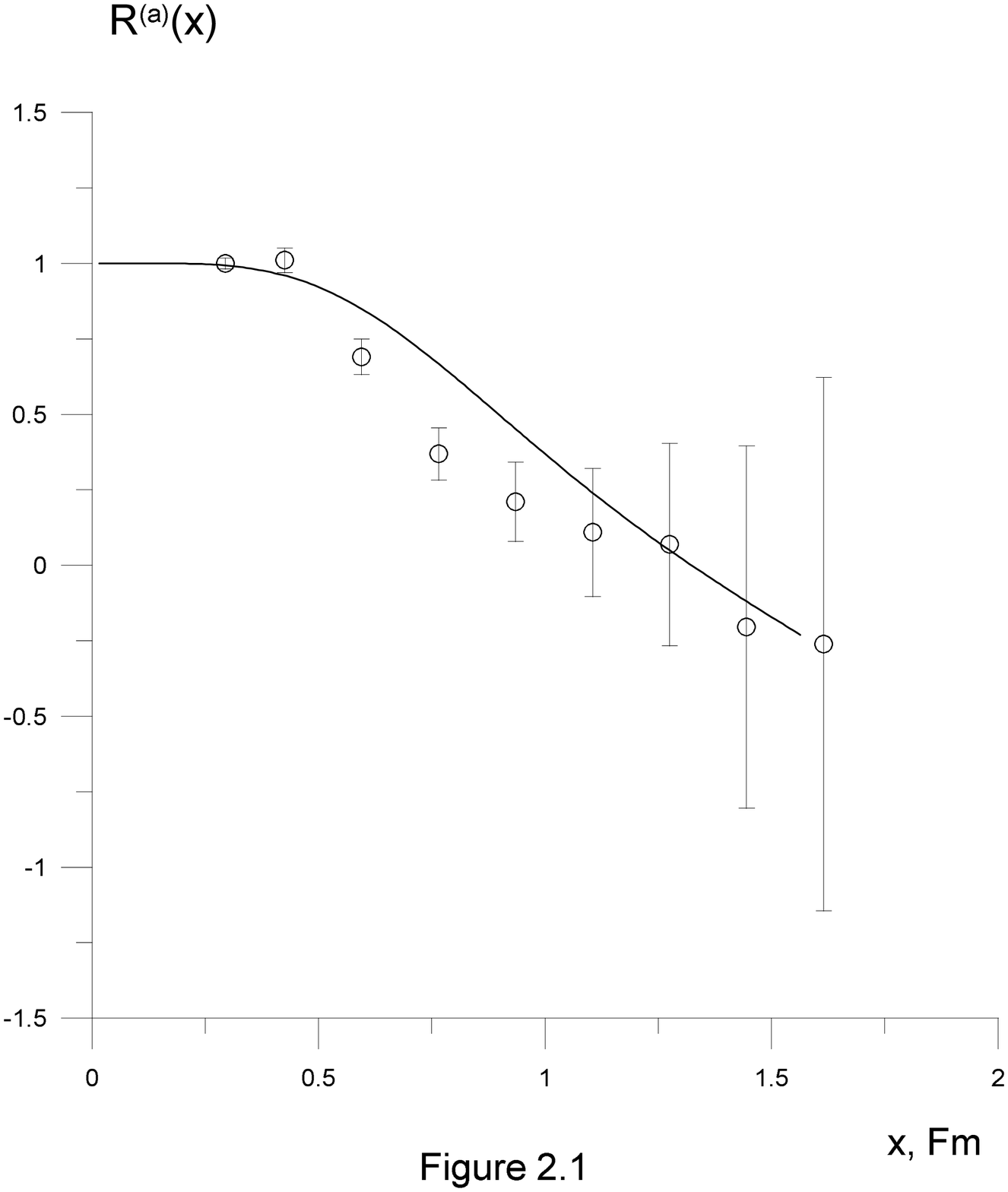}
&
\includegraphics[angle=-00,height=82mm,width=82mm,clip=true]{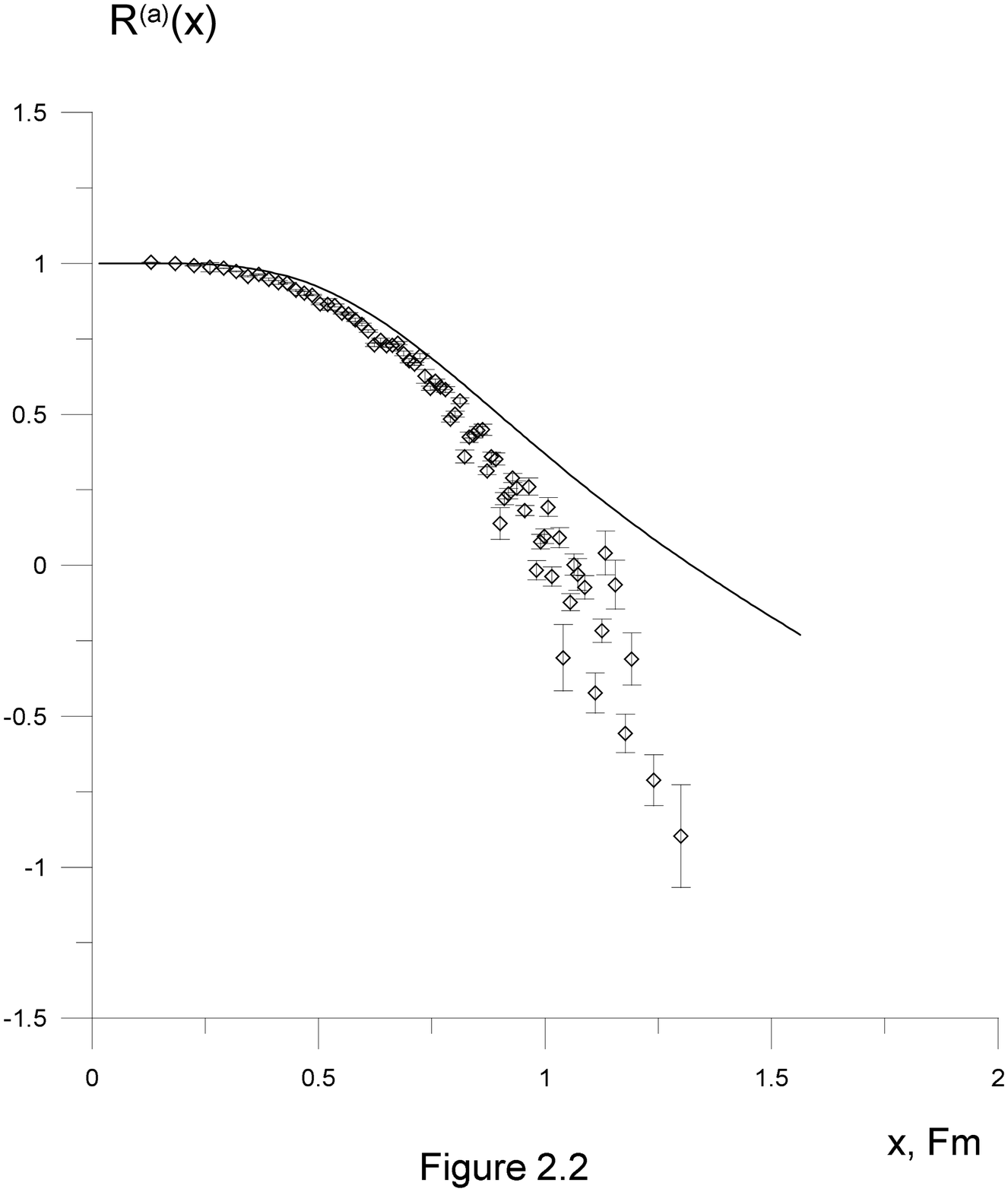}
\end{tabular}
\end{center}
\vspace{2cm}
\begin{center}
\includegraphics[angle=-00,height=82mm,width=82mm,clip=true]{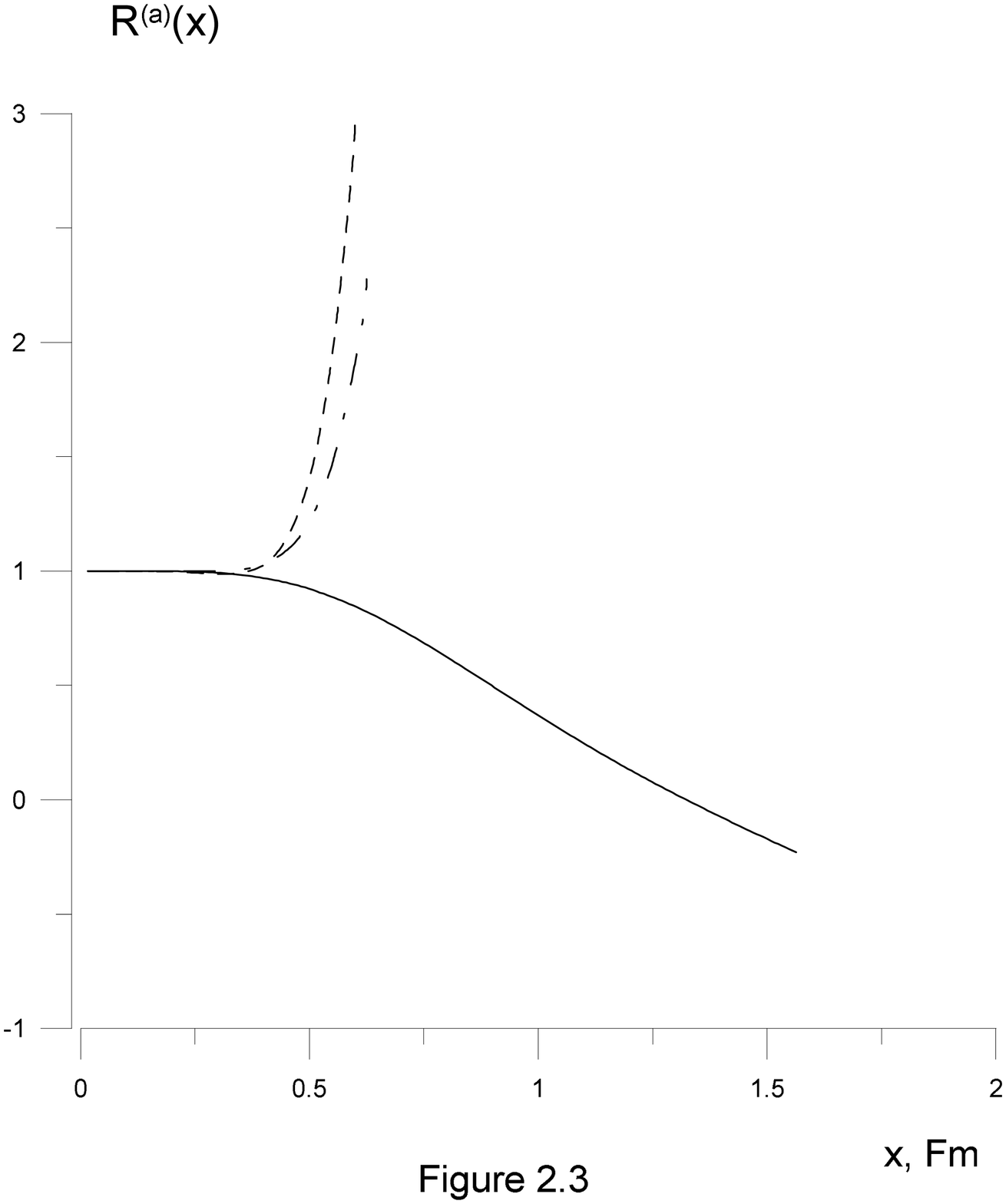}
\end{center}
\end{figure}

\begin{figure}[!htb]
\begin{center}
\begin{tabular}{lr}
\includegraphics[angle=-00,height=82mm,width=82mm,clip=true]{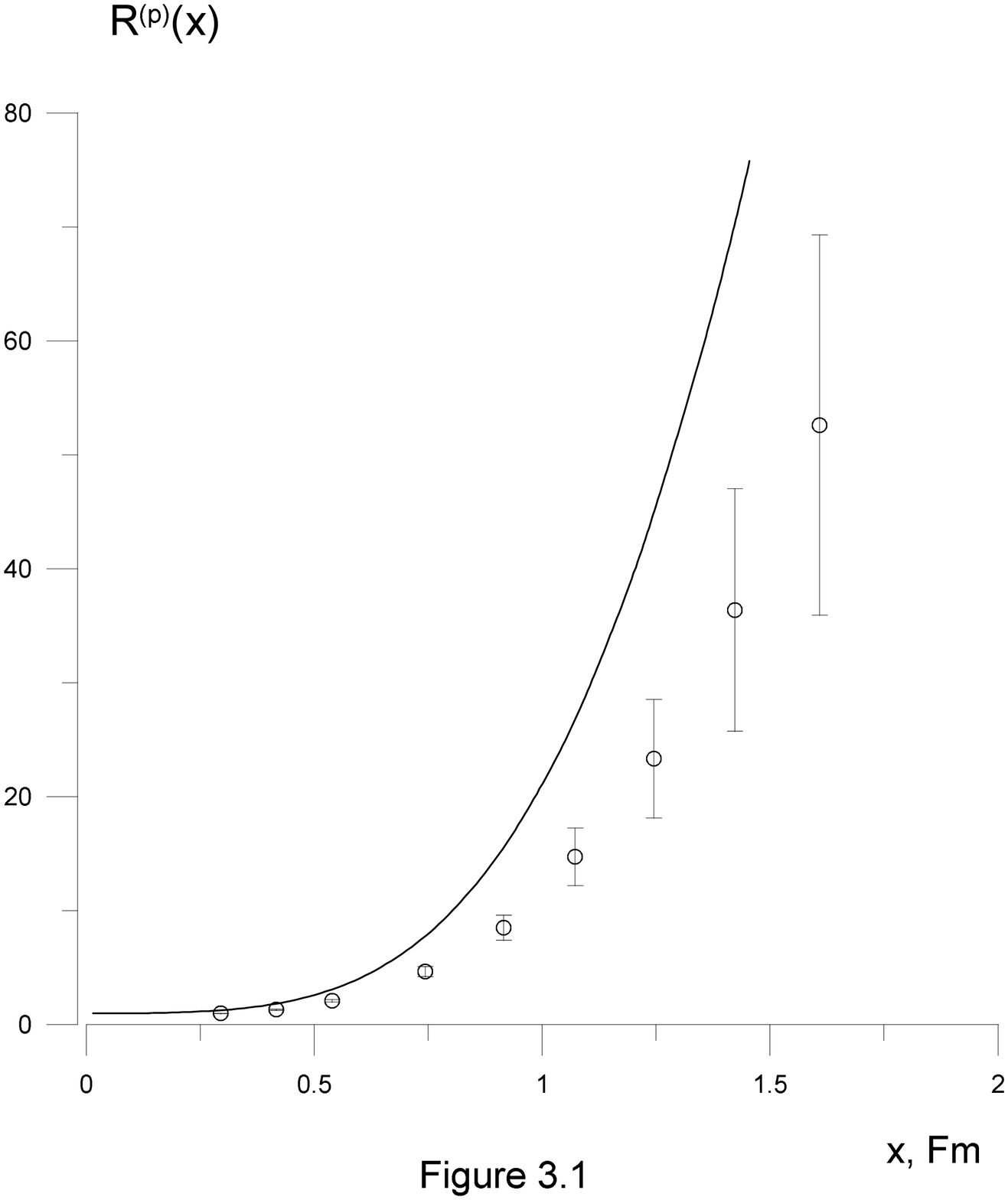}
&
\includegraphics[angle=-00,height=82mm,width=82mm,clip=true]{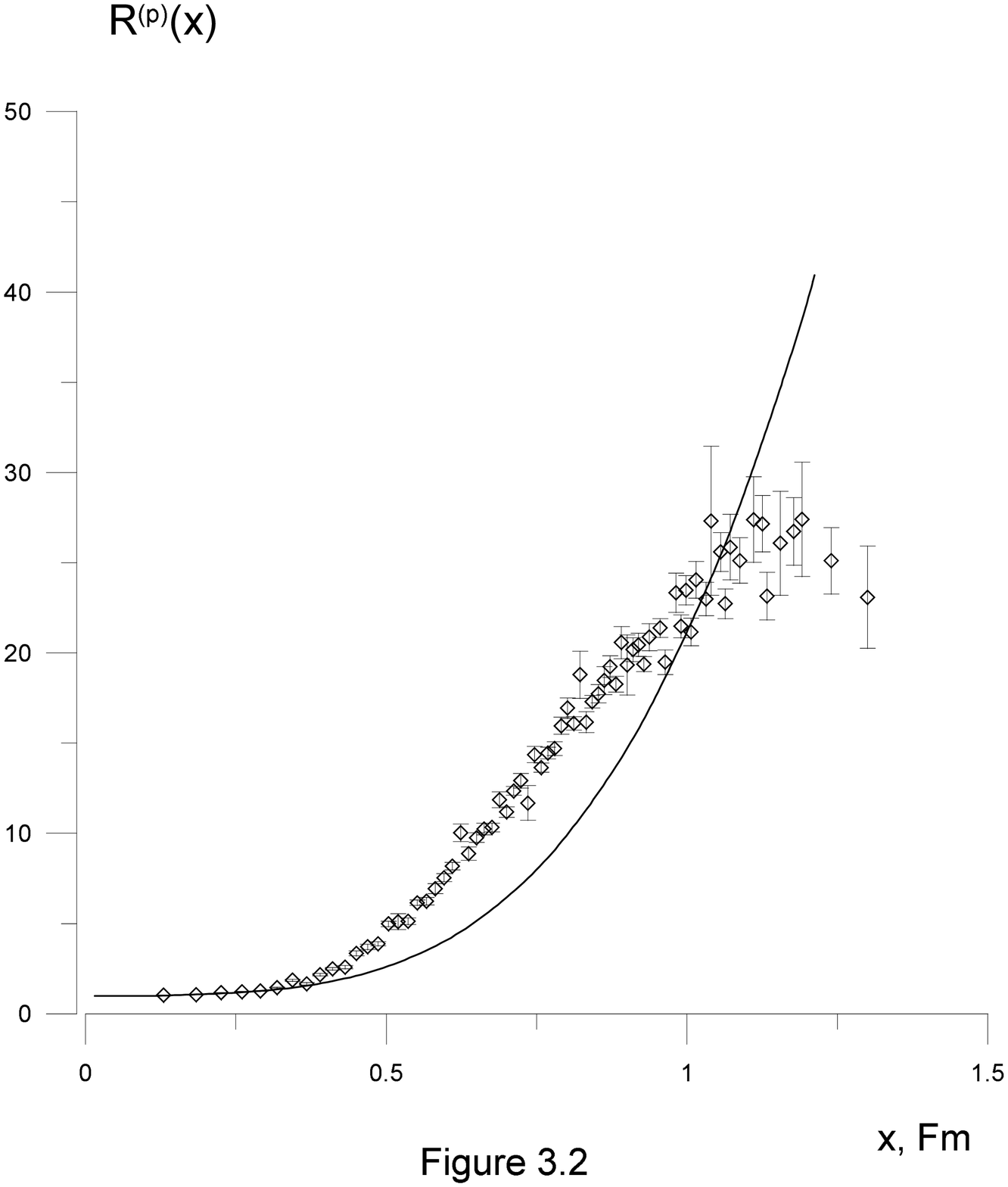}
\end{tabular}
\end{center}
\vspace{1.5cm}
\begin{center}
\begin{tabular}{lr}
\includegraphics[angle=-00,height=82mm,width=82mm,clip=true]{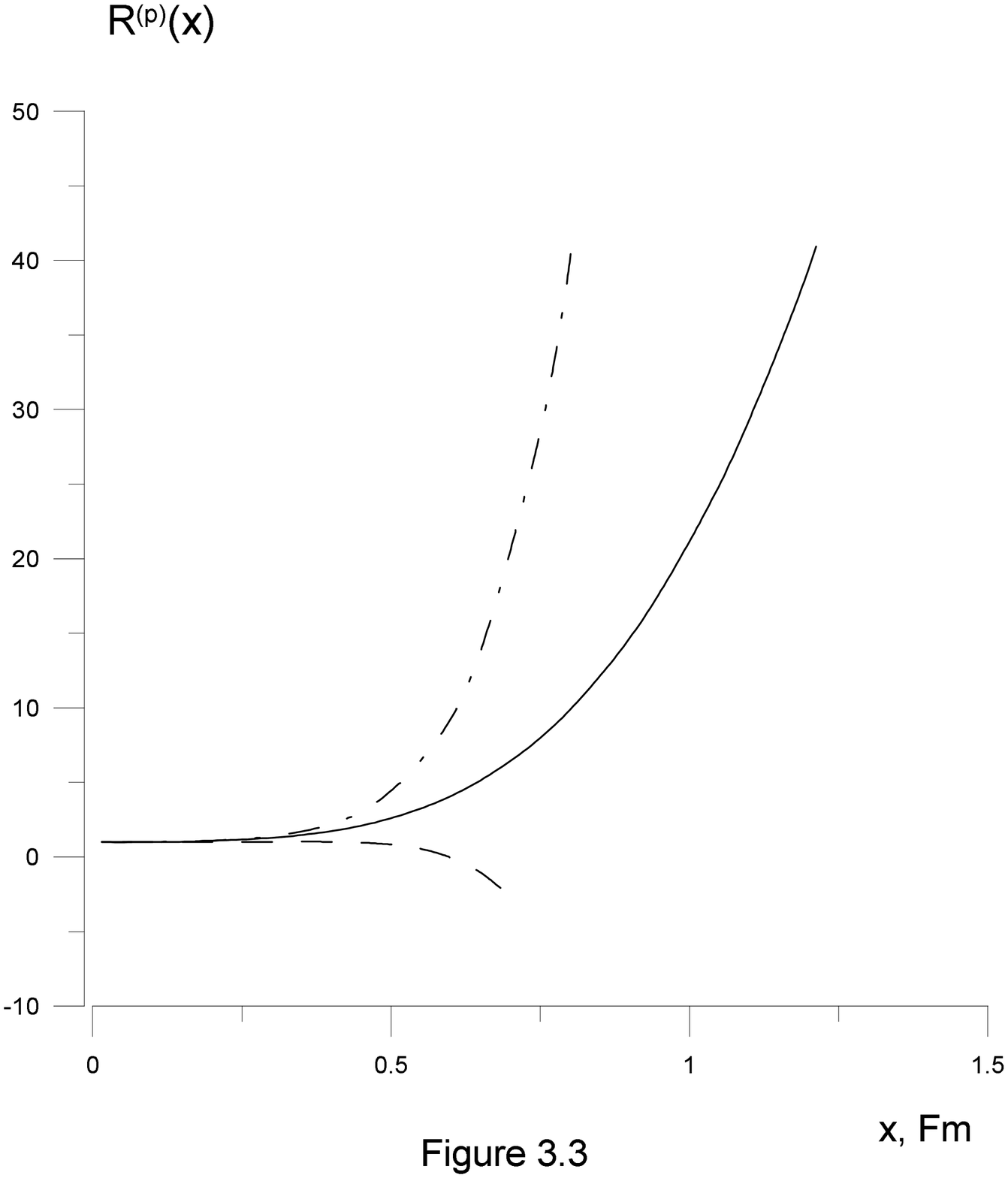}
&
\includegraphics[angle=-00,height=82mm,width=82mm,clip=true]{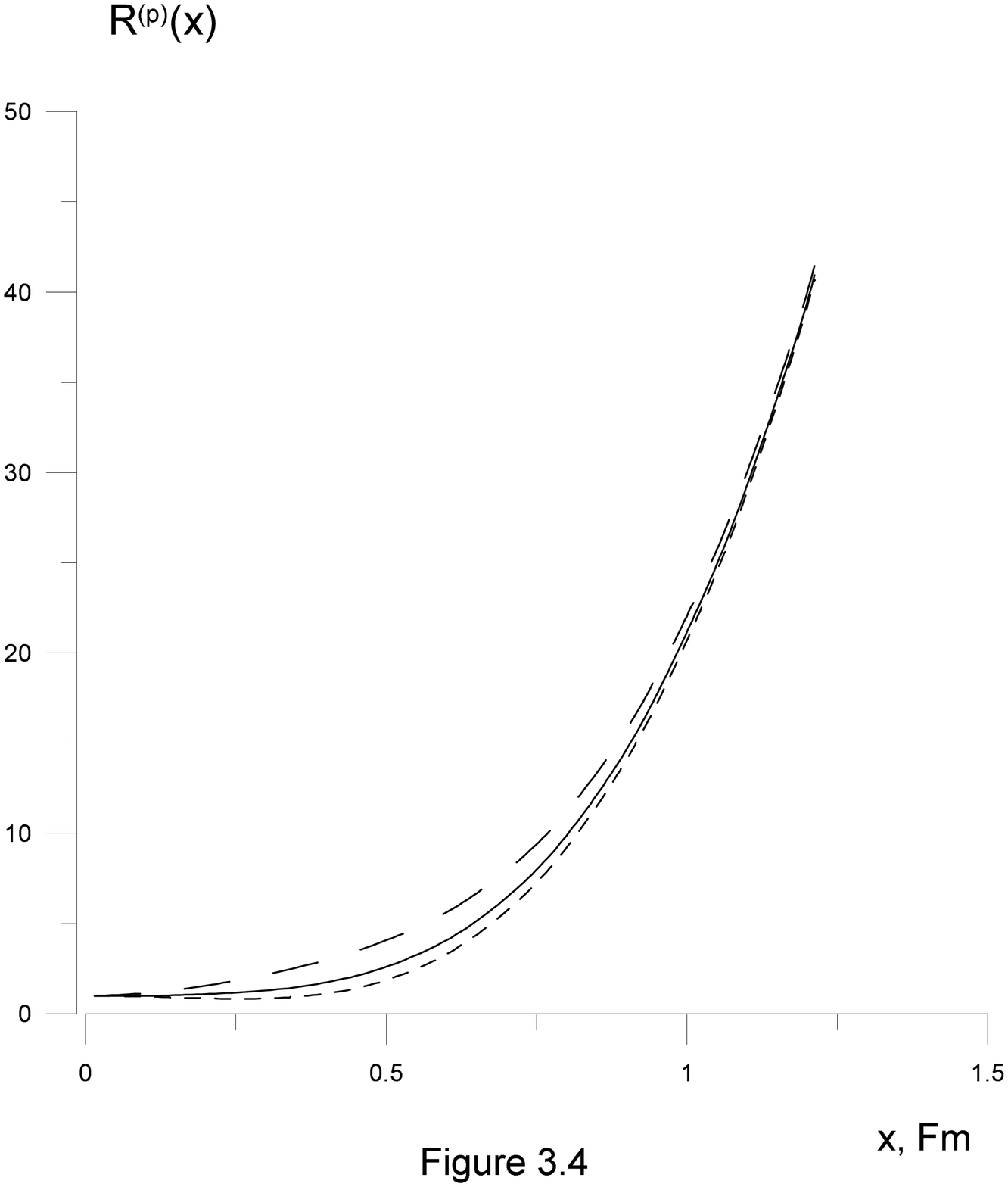}
\end{tabular}
\end{center}
\end{figure}

\begin{figure}[!htb]
\begin{center}
\begin{tabular}{lr}
\includegraphics[angle=-00,height=82mm,width=82mm,clip=true]{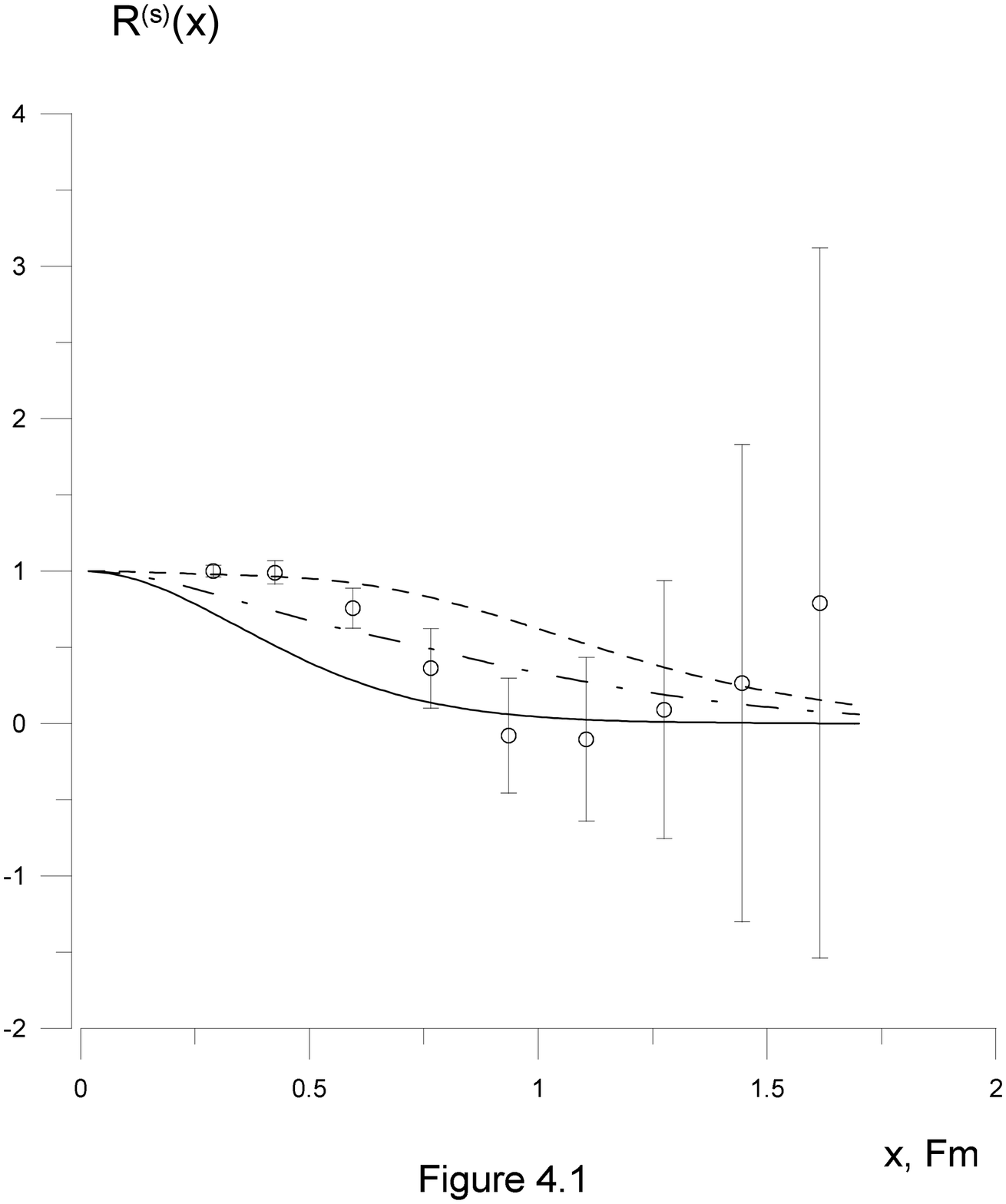}
&
\includegraphics[angle=-00,height=82mm,width=82mm,clip=true]{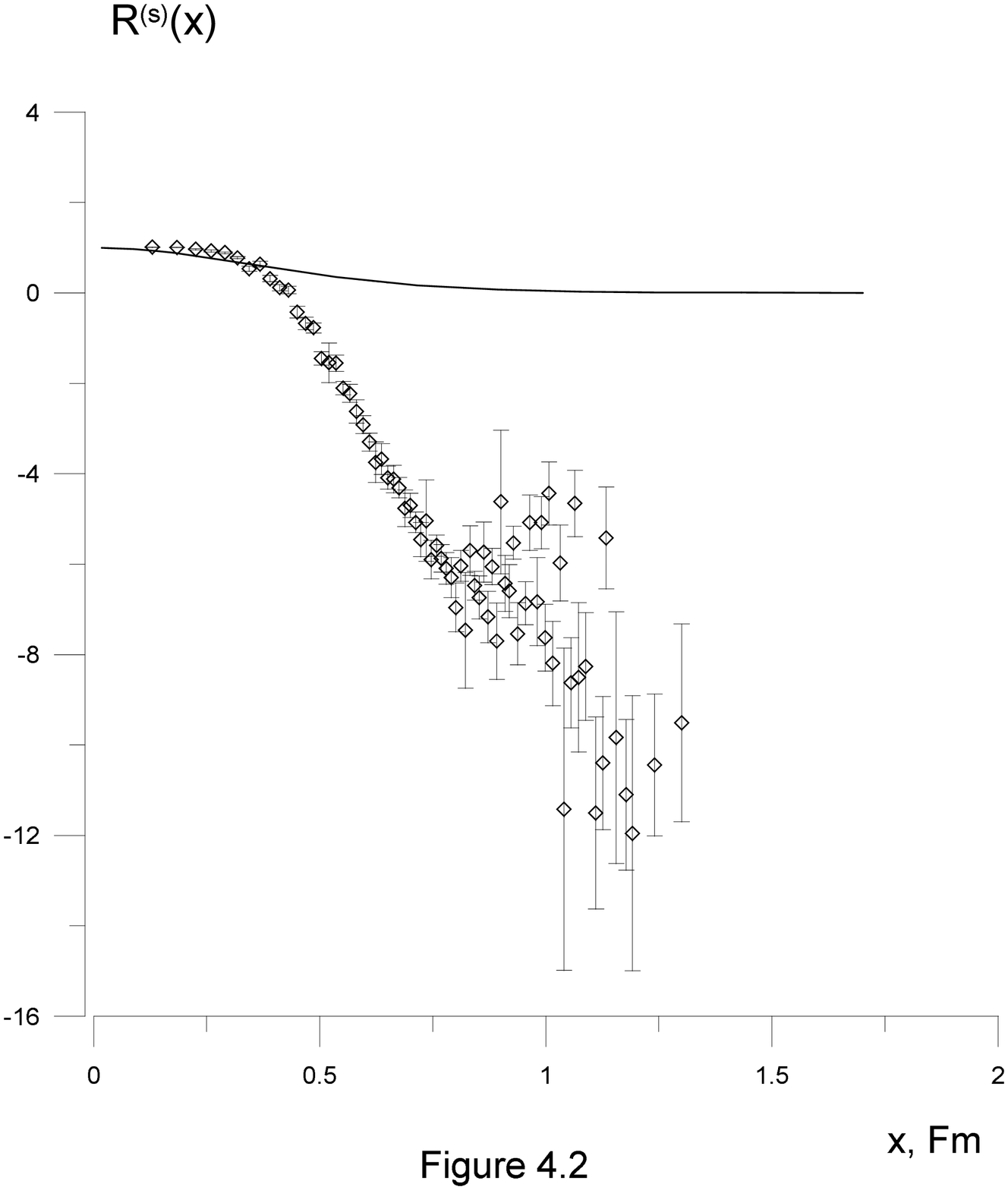}
\end{tabular}
\end{center}
\vspace{2cm}
\begin{center}
\includegraphics[angle=-00,height=82mm,width=82mm,clip=true]{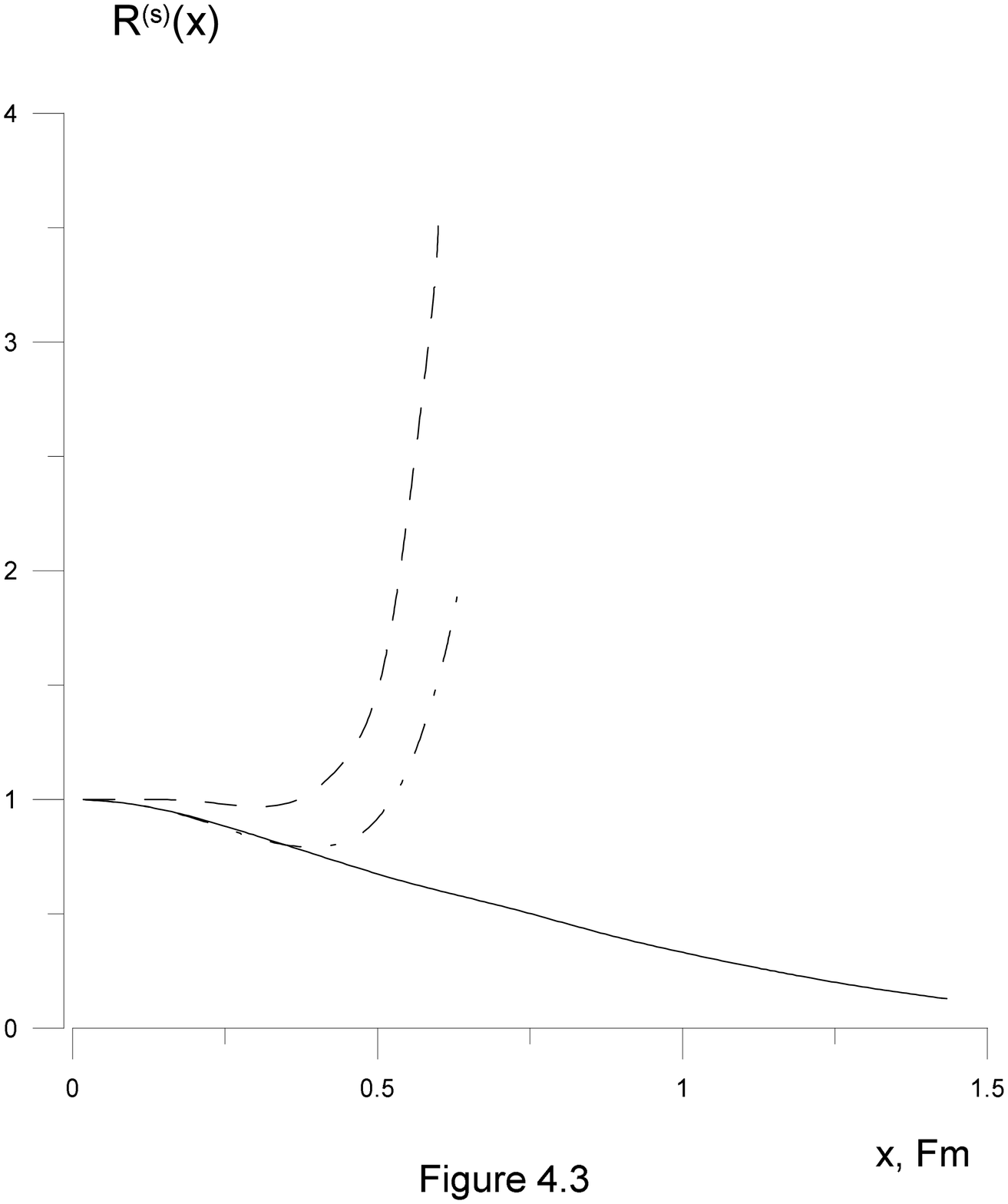}
\end{center}
\end{figure}

\end{document}